\documentclass[fleqn,twocolumn]{elsart}

\usepackage{graphics}
\usepackage{graphicx}
\usepackage{epsfig}

\usepackage{amssymb}

\begin{document}

\begin{frontmatter}



\title{Search for a Diffuse Flux of High-Energy 
Extraterrestrial Neutrinos with the NT200 Neutrino Telescope}


\author[a]{V. Aynutdinov},
\author[a]{V. Balkanov},
\author[d]{I. Belolaptikov},
\author[a]{L. Bezrukov},
\author[a]{D. Borschov},
\author[b]{N. Budnev},
\author[b]{A. Chensky},
\author[a]{I. Danilchenko},
\author[a]{Ya. Davidov},
\author[a]{G. Domogatsky},
\author[a]{A. Doroshenko},
\author[b]{A. Dyachok},
\author[a]{Zh.-A. Dzhilkibaev\corauthref{cor}},
\corauth[cor]{Corresponding author.}
\ead{djilkib@pcbai10.inr.ruhep.ru}
\author[f]{S. Fialkovsky},
\author[a]{O. Gaponenko},
\author[b]{O. Gress},
\author[b]{T. Gress},
\author[b]{O. Grishin},
\author[a]{A. Klabukov},
\author[h]{A. Klimov},
\author[a]{S. Klimushin},
\author[d]{K. Konischev},
\author[a]{A. Koshechkin},
\author[c]{L. Kuzmichev},
\author[f]{V. Kulepov},
\author[a]{B. Lubsandorzhiev},
\author[a]{S. Mikheyev},
\author[e]{T. Mikolajski},
\author[f]{M. Milenin},
\author[b]{R. Mirgazov},
\author[c]{E. Osipova},
\author[b]{A. Pavlov},
\author[b]{G. Pan'kov},
\author[b]{L. Pan'kov},
\author[a]{A. Panfilov},
\author[b]{Yu. Parfenov},
\author[a]{D. Petukhov},
\author[d]{E. Pliskovsky},
\author[a]{P. Pokhil},
\author[a]{V. Poleschuk},
\author[c]{E. Popova},
\author[c]{V. Prosin},
\author[g]{M. Rozanov},
\author[b]{V. Rubtzov},
\author[b]{Yu. Semeney},
\author[a]{B. Shaibonov},
\author[c]{A. Shirokov},
\author[e]{Ch. Spiering},
\author[b]{B. Tarashansky},
\author[d]{R. Vasiliev},
\author[a]{E. Vyatchin},
\author[e]{R. Wischnewski},
\author[c]{I. Yashin},
\author[a]{V. Zhukov}

\address[a]{Institute for Nuclear Research, 60th October Anniversary pr. 7a, 
Moscow 117312, Russia}
\address[b]{Irkutsk State University, Irkutsk, Russia}
\address[c]{Skobeltsyn Institute of Nuclear Physics  MSU, Moscow, Russia}
\address[d]{Joint Institute for Nuclear Research, Dubna, Russia}
\address[e]{DESY, Zeuthen, Germany}
\address[f]{Nizhni Novgorod State Technical University, Nizhni Novgorod, 
Russia}
\address[g]{St.Petersburg State Marine University, St.Petersburg, Russia}
\address[h]{Kurchatov Institute, Moscow, Russia}

\begin{abstract}
We present the results of a search for high energy extraterrestrial neutrinos
with the Baikal underwater Cherenkov detector {\it NT200}, 
based on data taken in 1998 - 2003.
Upper limits on the diffuse fluxes of
$\nu_e+\nu_{\mu}+\nu_{\tau}$, predicted by several models 
of AGN-like neutrino sources, are derived.  
For  an $E^{-2}$ behavior of the neutrino spectrum,
our limit is
 $E^2 \Phi_{\nu}(E)<8.1\times 10^{-7}\,
\mbox{cm}^{-2}\,\mbox{s}^{-1}\,
\mbox{sr}^{-1}\,\mbox{GeV}$ 
over an neutrino energy range $2\times10^4 \div 5 \times 10^7\,\mbox{GeV}$.
The upper limit on the resonant $\bar{\nu}_e$ diffuse flux is 
$\Phi_{\bar{\nu}_e}<$3.3$\times$10$^{-20}$ 
cm$^{-2}$s$^{-1}$sr$^{-1}$GeV$^{-1}$.
\end{abstract}

\end{frontmatter}

\section{Introduction}
\label{s1}

High energy neutrinos are likely produced in many violent processes 
in the Universe. Their detection would unambiguously reveal the hadronic
nature of the underlying processes. Neutrinos would be generated by 
proton-proton or proton-photon interactions followed by  production 
and decay of charged mesons. 

One detection mode of neutrino telescopes is the identification of 
individual, point-like sources of high energy neutrinos. Galactic candidates 
for these objects include supernova remnants or microquasars, extragalactic 
objects are e.g. active galactic nuclei (AGN) and Gamma Ray Bursts (GRB). 
Individual sources might be too weak to produce an unambiguous directional 
signal, however the integrated neutrino flux from all sources 
 could produce a detectable diffuse neutrino signal. 
Astrophysical neutrinos generated in top-down models are,
by definition, of diffuse nature.
A diffuse neutrino flux can be identified by a high-energy 
excess on top of the background of  known fluxes of charged particles 
recorded by a  neutrino telescope. Such charged particles are dominantly muons 
produced in the atmosphere above the detector, with a small contribution
from  muons generated in interactions of atmospheric neutrinos. 

In this paper we present results of a search for neutrinos with energies 
larger than 10 TeV.  The analysis is based on data taken with the Baikal 
neutrino telescope {\it NT200} between April 1998 and February 2003.
Instead of focusing to high-energy particles crossing the array,  the  
analysis is tailored to signatures of isolated high-energy cascades 
in a large volume around the detector \cite{APP3}. This search strategy 
dramatically enhances the sensitivity of {\it NT200} to diffuse high energy processes.

The cascades can stem from leptons and hadrons produced in high energy 
charged current processes
\begin{equation}
\nu_l(\bar{\nu_l}) + N \stackrel{CC}{\longrightarrow} l^-(l^+) + 
\mbox{hadrons},
\end{equation}
or from the hadronic vertex of neutral current processes
\begin{equation}
\nu_l(\bar{\nu_l}) + N \stackrel{NC}{\longrightarrow} 
\nu_l(\bar{\nu_l}) + \mbox{hadrons},
\end{equation}
where $l = e, \mu$ or $\tau$. 
The energy released by the hadronic 
cascade in reaction (2) is small compared to that of the leptonic cascades 
in (1). Since only electrons and taus develop cascades (electrons by directly 
showering up, taus via their decay to secondary particles which develop 
a cascade), the sensitivity of this search is dominated by  $\nu_e$ and 
$\nu_{\tau}$ detection.

Cascades can also be produced by resonant W production
\begin{equation}
\bar{\nu_e} + e^- \rightarrow W^- \rightarrow \mbox{anything},
\end{equation}
with the resonant neutrino energy  
$E_0=M^{2}_w/2m_e=6.3\times 10^6 \,$GeV 
and cross section $5.02\times 10^{-31}$cm$^2$.

This paper is organized as follows: In section 2 we describe detector and 
site, and illustrate the detector performance. In section 3 we first line 
out the search strategy. This is followed by a description of how a laser 
light source mimicking high energy cascades is used to investigate the 
detector performance with respect to energetic cascades. Furthermore, we 
describe the simulation of signal and background processes. Section 4 is 
devoted to the analysis of the experimental data. Limits on high energy 
neutrino fluxes and comparison to models are presented in section 5. 
Section 6 gives conclusions and a short outlook on how to further improve 
the present limits.

\section{Detector and site}

\begin{figure*}[htb]
 \begin{center}
\includegraphics[width=8.3cm,height=9.cm,angle=-90]{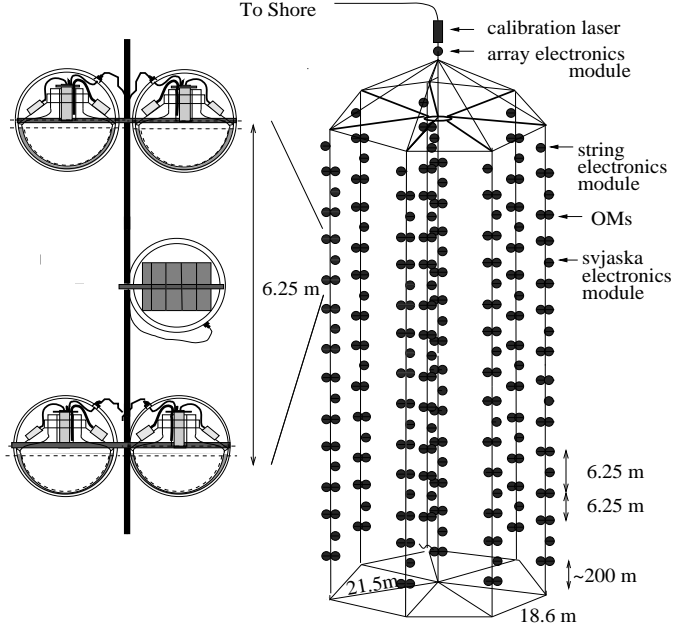}
\end{center}
\caption{
A schematic view of {\it NT200}. The expansion left-hand 
shows 2 pairs of optical modules (``svjaska'') with the svjaska
electronics module, which houses part of the readout and control electronics.
}
\label{fig1}
\end{figure*}

The Baikal Neutrino Telescope  is operated in Lake 
Baikal, Siberia,  at a depth of \mbox{1.1 km}. 
The present stage of the telescope,
{\it NT200} \cite{APP1,APP2,Bal1}, 
takes data since
April 6th, 1998 and consists of 192
optical modules (OMs). A schematic view of the Baikal Telescope {\it NT200}
is shown in fig. \ref{fig1}. 
An umbrella-like frame carries  8 strings,
each with 24 pairwise arranged OMs.
All OMs face downward, with the exception of 
the second and eleventh pairs on each string which face
upward. Three underwater electrical cables 
connect the detector with the shore station. 
Each OM contains a 37-cm diameter {\it QUASAR} - photo multiplier (PM), 
which has been developed specially for our project \cite{OM2}. 
The PMs record the Cherenkov light produced by charged particles
in water. The two PMs of a pair are switched in coincidence in order to 
suppress background from bioluminescence and PM noise. 
A pair of OMs defines a {\it channel}. 
The light arrival time assigned to a channel is the response
time of the OM with the earliest hit. The amplitude assigned
to a channel is that recorded by one pre-selected PM of the two
PMs in a pair. For those channels for which one of the two OMs failed
and the single PM noise rates of the remaining OM are not too
high (smaller than the average noise rate of 100 kHz), that OM
is operated in single mode (1PM/channel), with
a threshold slightly higher than the typical 0.3 photo electrons.

A {\it trigger}
is formed by the requirement of \mbox{$\geq N$ {\it hits}}
(with {\it hit} referring to a channel) within \mbox{500 ns}.
$N$ is typically set to 
\mbox{3 or 4.} For  these events, amplitude and time of all fired
channels are digitized and sent to shore. 

\begin{figure*}[htb]
\includegraphics[width=.52\textwidth,height=5.2cm]{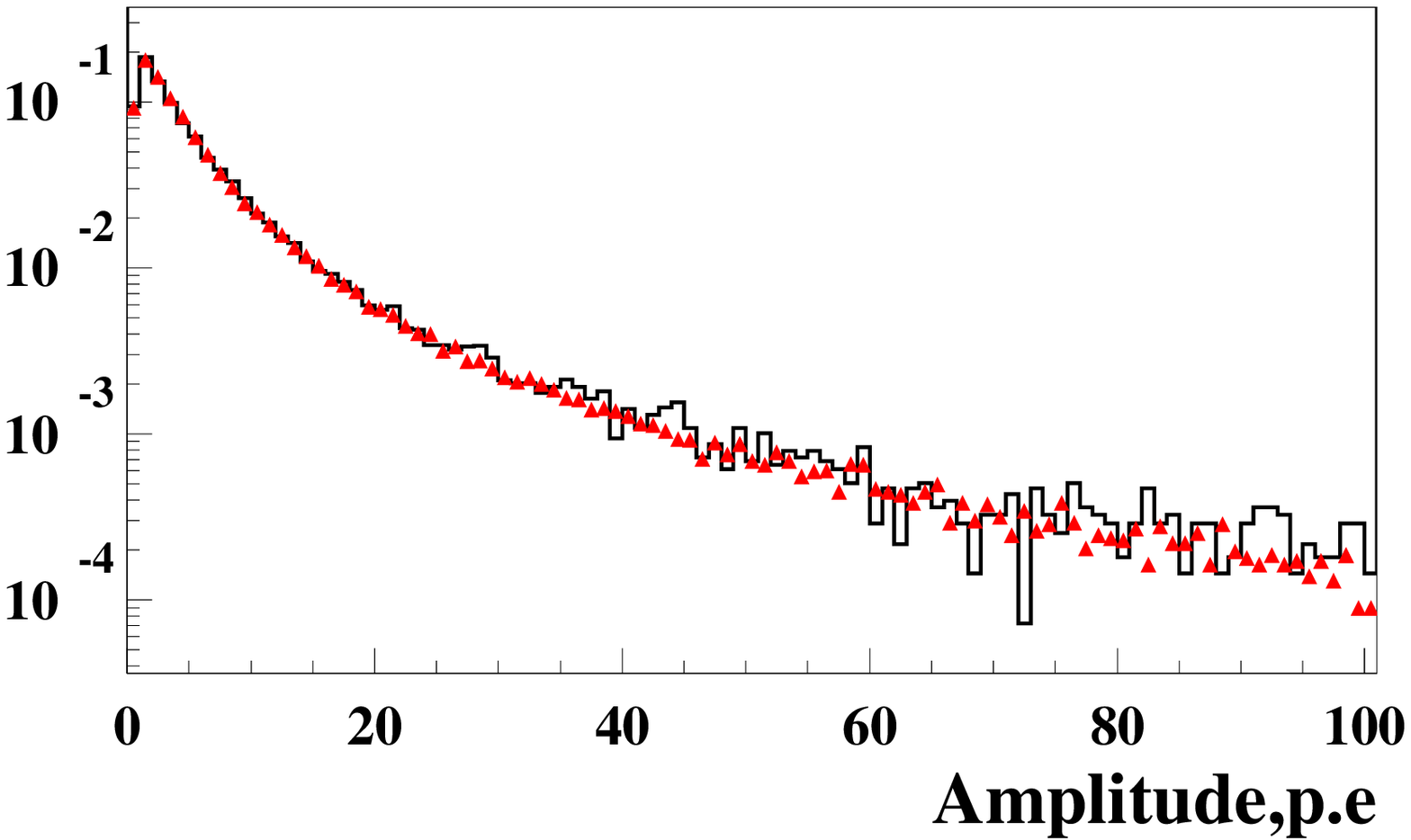}
\hfill
\includegraphics[width=.52\textwidth,height=5.2cm]{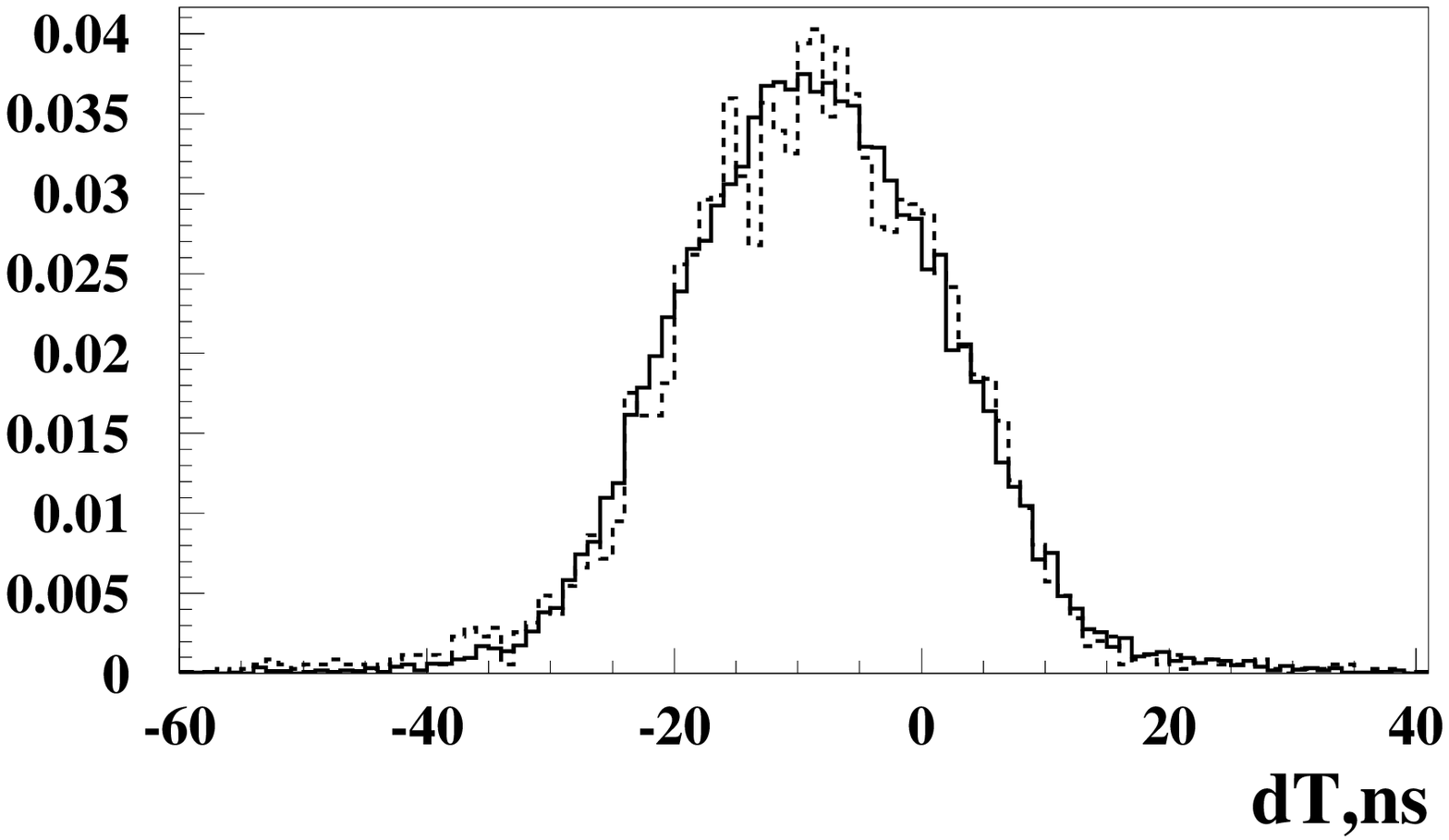}
\\
\caption{Left panel:
the normalized amplitude distribution for channel 2
(points -- experiment, histogram -- 
atmospheric muon simulation).
Right panel: the normalized
time difference of channel 42 and channel 43. Histogram - experiment,
dashed histogram - expectation from atmospheric muons.
}
\label{fig2}
\end{figure*}

Two nitrogen lasers are used for calibration of the detector.
The first one ({\it fiber laser})
is mounted just above the array. Its light is guided via optical
fibers of equal length to each OM pair. The fiber laser provides
the OMs with simultaneous light signals in order to determine the offset
for each channel. The second laser ({\it water laser}) is arranged 20 m
below the array. Its light propagates through the water.
This laser serves to monitor the water quality, in addition to
dedicated environmental devices located along a separate string.
In the context of this analysis, however, its main  purpose
is to simulate high energy particle cascades outside the geometrical
volume of {\it NT200}. The maximum light intensity of the water laser 
($\sim$10$^{11}$ photons/pulse) roughly corresponds to the Cherenkov 
radiation emitted by high-energy cascades with $E_{cas}\approx$1 PeV. 
A full cycle of detector calibration running both lasers over a wide range of
intensities is repeated every third day.

Figure 2 demonstrates the level to which basic features of the  detector are
understood and can be reproduced by Monte-Carlo (MC) simulations. We have used
atmospheric muons  as a high statistics standard calibration signal.
The figure shows distributions of recorded amplitudes and light arrival
time differences compared to MC-simulation.
The experimental data are in good agreement with
simulation (see also \cite{APP1}).

Lake Baikal deep water is characterized by an
absorption length of $L_{abs}$(480 nm)=20$\div $24 m, a scattering
length of $L_s=$30$\div $70 m and a strongly anisotropic scattering
function $f(\theta)$ with a mean cosine of the scattering angle 
$\overline{\cos}(\theta)=0.85 \div 0.9$.

\section{The analysis method}

\subsection{Search strategy}
\begin{figure}
\begin{center}
\includegraphics*[width=.5\textwidth,height=6.0cm]{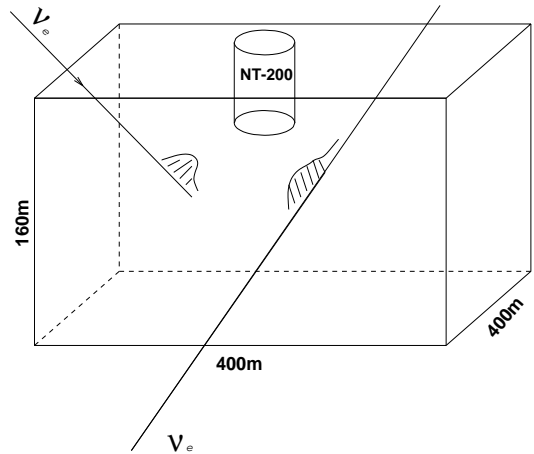}
\end{center}
\caption{Detection principle for neutrino induced cascades
with {\it NT200}.}
\label{fig4}
\end{figure}

The BAIKAL survey for high energy neutrinos searches
for bright cascades produced at the neutrino interaction
vertex in a large volume around 
and below
the neutrino telescope (see fig. \ref{fig4}).
Lack of significant light scattering allows to monitor a 
volume exceeding the geometrical volume by more than an order of magnitude.
This results in sensitivities 
for high energy cascade detection which are close
to those of the much larger AMANDA detector \cite{AMANDAHE}. 
The main background source to this analysis are atmospheric muons,
with a flux 10$^6$ times higher than that of atmospheric neutrinos.

We select events with high multiplicity of hit channels $N_{\mbox{\small hit}}$,
corresponding to bright cascades. 
The main remaining background to isolated cascades from neutrino
interactions are then cascades from bremsstrahlung along energetic
downward muons. To separate high-energy neutrino events
from background events a cut on the variable 
$t_{\mbox{\footnotesize min}}=\mbox{\footnotesize min}(t_i-t_j)$ 
(with $i<j$) is applied.
Here, $t_i, \, t_j$ are the arrival times at channels $i,j$ 
on each string (the numbering of channels is from top to bottom along strings),
the minimum over all strings is calculated.
Positive and negative values of $t_{\mbox{\footnotesize min}}$ correspond to 
upward and downward propagation of light, respectively. We require
\begin{equation}
\label{eq1}
t_{\mbox{\footnotesize min}}>-10 \,\, \mbox{ns}.
\end{equation}
This cut accepts only time patterns corresponding to upward traveling 
light signals. It rejects most events from brems-cascades produced by
downward going muons since the majority of muons is close to 
the vertical; they would  cross the detector or pass nearby and generate 
a downward time pattern. Only few muons with large zenith angles may 
escape this cut and illuminate the array by their own Cherenkov radiation or 
that from  bright cascades from below. 

\begin{figure}
\begin{center}
\includegraphics*[width=.45\textwidth,height=6.0cm]{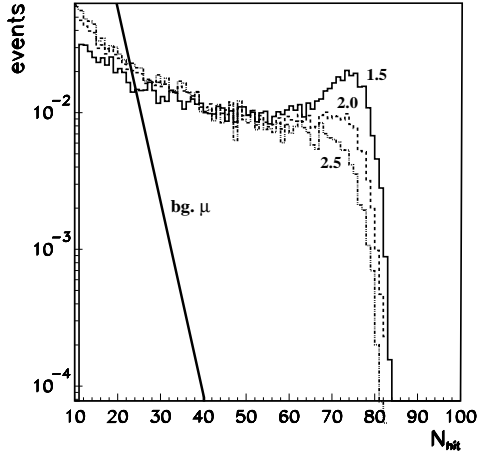}
\end{center}
\caption{The normalized hit multiplicity distributions of events
from $\nu_e$ fluxes after
selection criterion (\ref{eq1}). Solid, dashed and dotted histograms correspond
to $\gamma$=1.5, 2, 2.5, respectively. Also shown is the 
$N_{\mbox{\small hit}}$ distribution of background events from atmospheric 
muons
(thick line).
}
\label{fig5}
\end{figure}

The energy spectrum of 
neutrinos from galactic and cosmological sources
or from the decay of topological defects is expected to have
a significantly flatter shape than the spectrum of 
atmospheric muons. This is reflected by different 
$N_{\mbox{\small hit}}$ distributions.
In fig. \ref{fig5} we show $N_{\mbox{\small hit}}$ distributions of 
simulated events which survive cut (\ref{eq1}) and would be induced by electron 
neutrinos with fluxes of shape $\sim E^{-\gamma}$, 
with \mbox{$\gamma$=1.5, 2 and 2.5}
(normalized to each other, simulations is for 80 operating channels). 
Also shown is the $N_{\mbox{\small hit}}$ distribution of background events 
induced by atmospheric muons (normalized arbitrarily to neutrino events). 
This distribution is much steeper, so that an extraterrestrial neutrino 
signal would appear
as an excess of events with large $N_{\mbox{\small hit}}$ above the
background of atmospheric muons.

\subsection{Laser calibration}
\begin{figure*}
\includegraphics*[width=.5\textwidth,height=6.0cm]{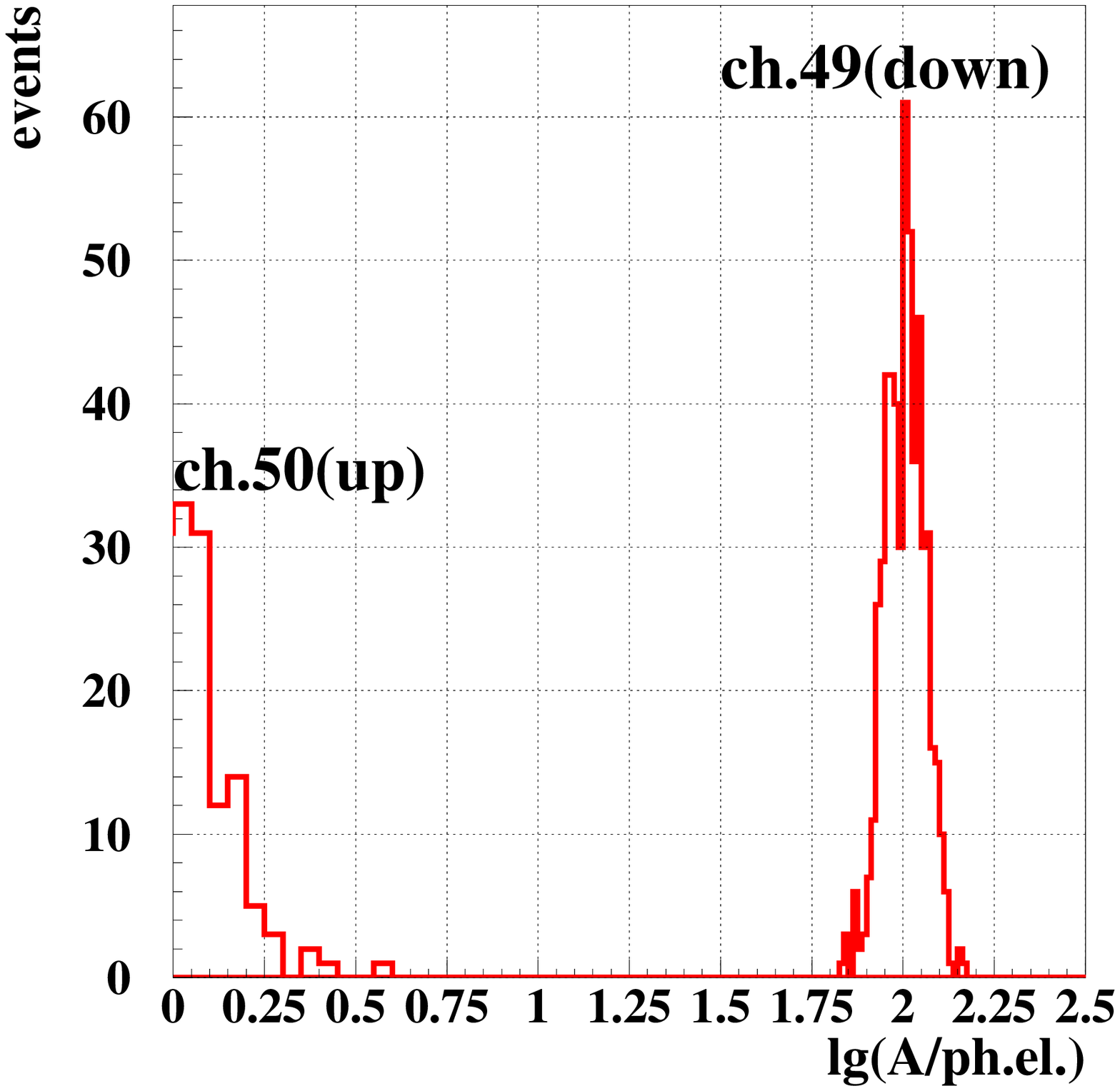}
\hfill
\includegraphics*[width=.5\textwidth,height=6.0cm]{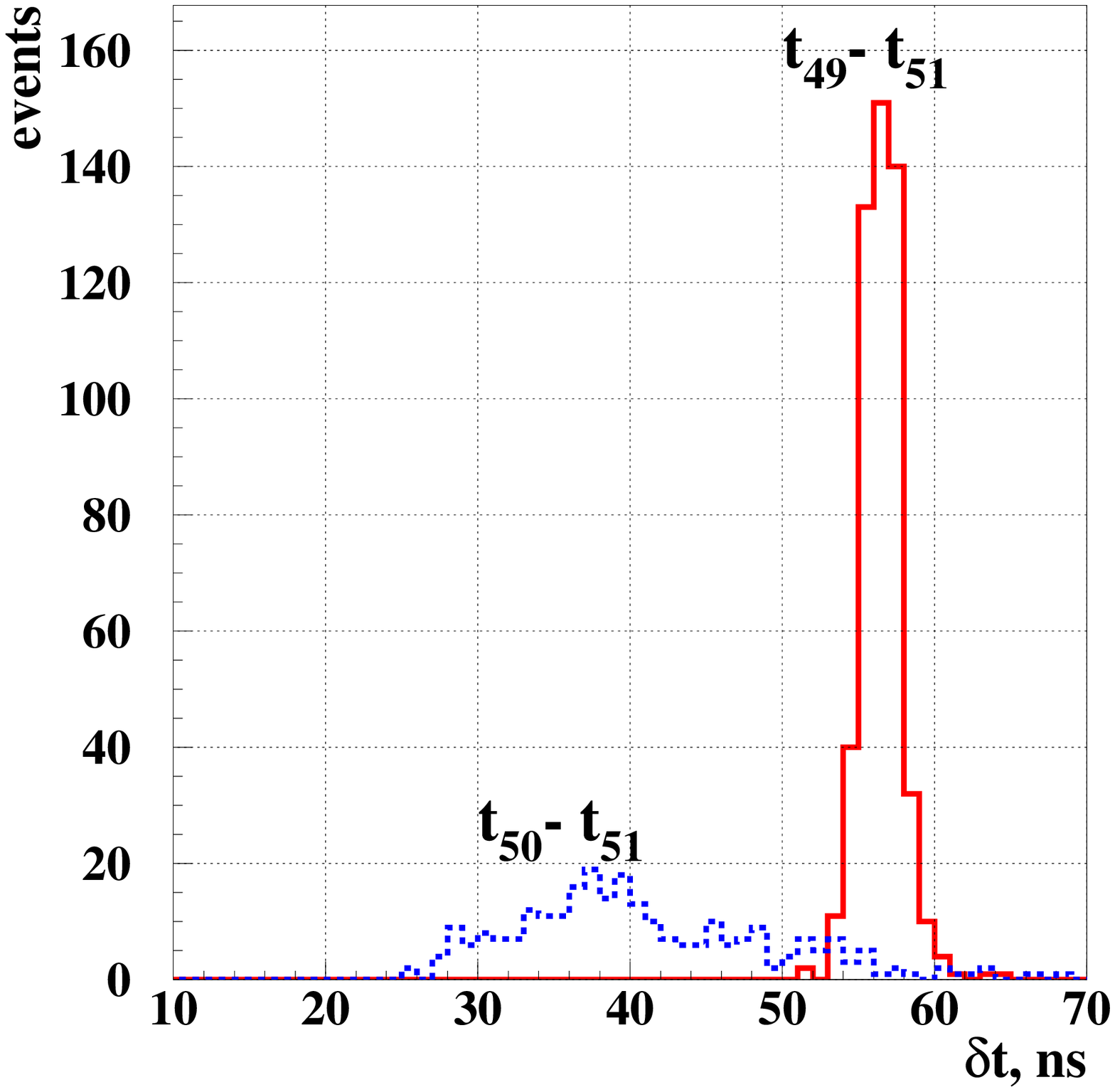}
\\
\caption{Laser calibration: left panel -  the amplitude distributions of channels directed
 downward (channel 49) and
upward (channel 50). Right panel: time difference 
distribution $\delta t$ for pairs of channels facing both 
downward (49,51) and upward (50)/downward (51),
respectively.}
\label{fig6}
\end{figure*}
As mentioned above, a central purpose of the water laser
is to investigate the response of {\it NT200} to bright
light sources below the detector.
The water laser is mounted on the central string about 20 m 
below the bottom OMs of {\it NT200}. An attenuator allows to 
operate the device with five gradually decreasing light pulse intensities.
In the most powerful mode, about 10$^{11}$ photons/pulse are emitted
isotropically.
Figure \ref{fig6} (left panel) shows the amplitudes
on channel 49 (turned downward) and channel 50 (turned upward)
induced by these pulses. The distances between the water laser
and channels 49 and 50 
(both on central string and above water laser)
are 88.8 m and 81.3 m, respectively.
In contrast to channel 49, which faces the light source,
channel 50 is directed in the opposite direction.
Therefore it is almost blind to direct light and detects mostly scattered 
photons. Due to the low scattering coefficient of Baikal water, 
the signals in channel 50 are about 70 times lower than in  channel 49.

\begin{figure*}
\includegraphics*[width=.5\textwidth,height=6.0cm]{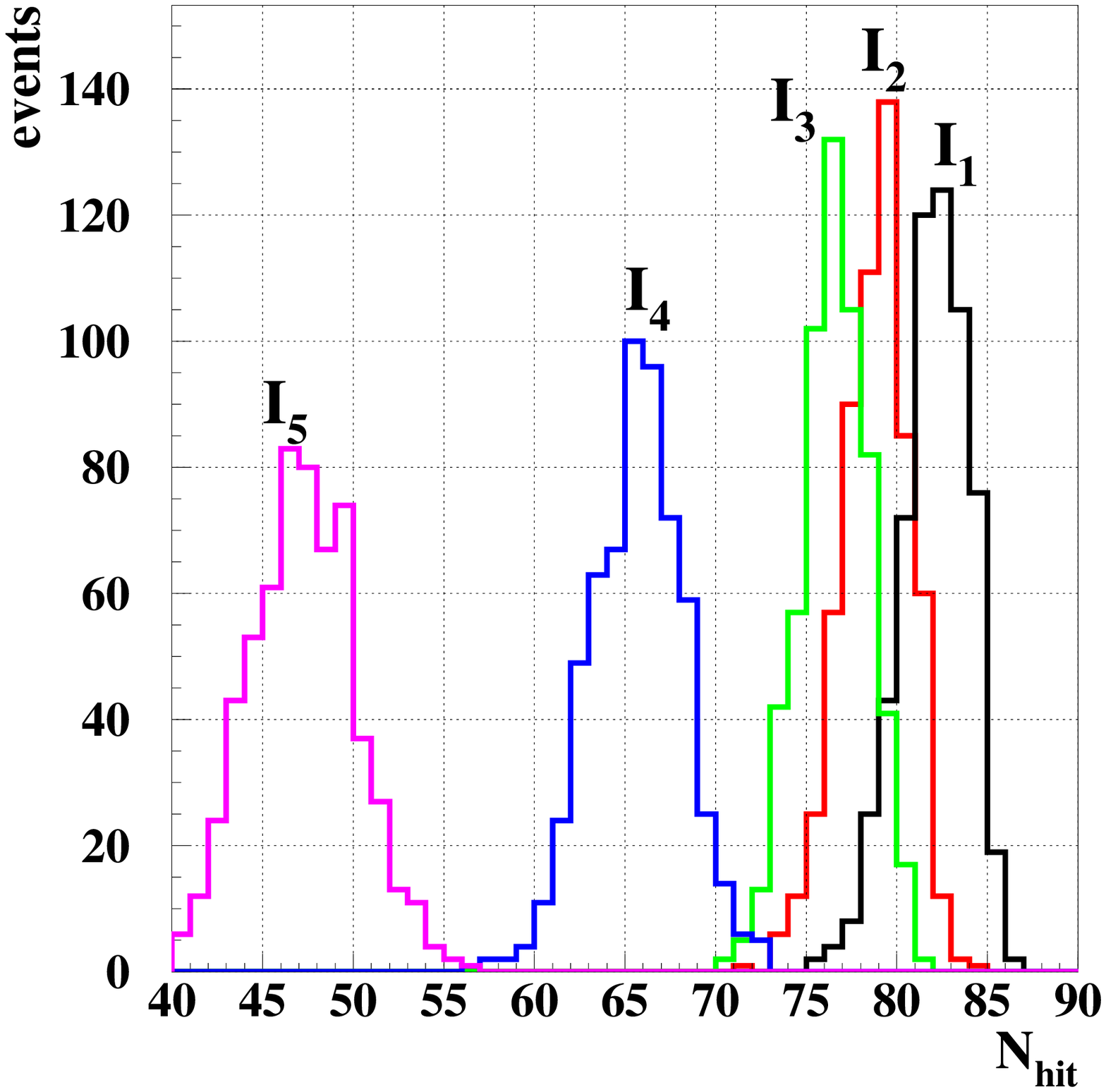}
\hfill
\includegraphics*[width=.5\textwidth,height=6.0cm]{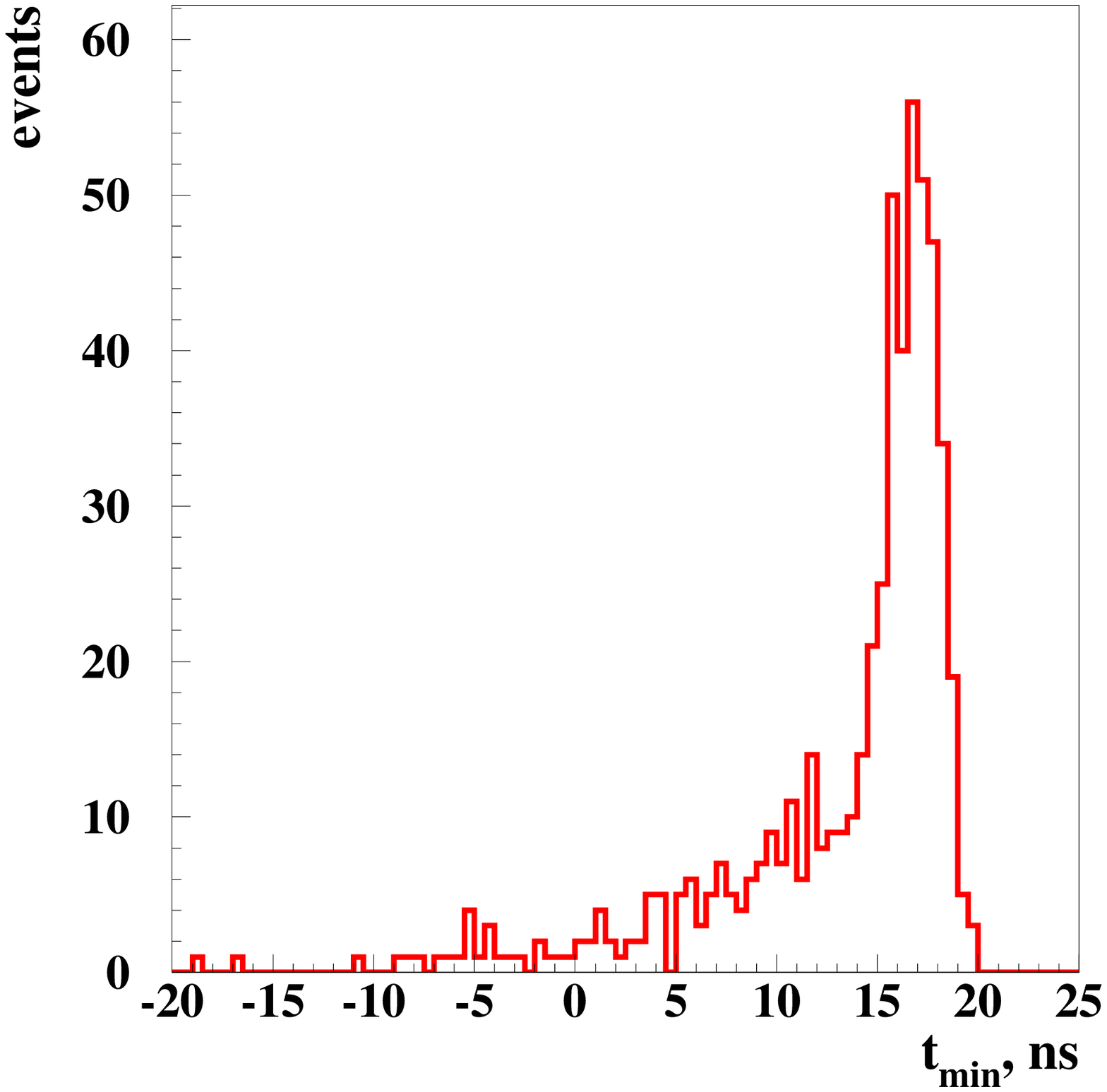}
\\
\caption{Left panel: the hit channels multiplicity of events induced by
laser pulses with five different intensities for.
Right panel: the $t_{\mbox{\footnotesize min}}$ distribution for laser events.
}
\label{fig7}
\end{figure*}

Figure \ref{fig6} (right panel) shows the difference $\delta t$ of laser 
light arrival times on channel 49 (channel 50) and 
channel 51. Like channel 49, channel 51 is turned downward.
It is located 12.5 m below channel 49. The r.m.s. of
$\delta t=(t_{49}-t_{51})$, i.e. for the two downward looking channels,
is about 1.4 ns. It is not affected by light scattering in water 
and entirely due to experimental time resolution.
The r.m.s. of $\delta t=(t_{50}-t_{51})$ is about 8.4 ns.
This large value is caused by the dispersion of photon arrival times
on channel 50 due to light scattering.
To reach maximum background suppression in the cascade search by
application of strict arrival time cuts, we exclude upward looking
channels from the analysis.

The distributions of the variables $N_{\mbox{\small hit}}$ and 
$t_{\mbox{\footnotesize min}}$,
which are used for the selection of neutrino induced events,
are shown in fig. \ref{fig7}.
Distributions on the left panel correspond to
the five laser intensities - 1.2$\times$10$^{11}$,
2.2$\times$10$^{10}$, 4.7$\times$10$^{9}$, 1.0$\times$10$^{9}$ and 
2.5$\times$10$^{8}$ photons/pulse -  which 
approximately correspond
to cascade energies 1200, 220, 47, 10 and 2.5 TeV, respectively.
The $t_{\mbox{\footnotesize min}}$ distribution on the right panel corresponds
to the most powerful laser intensity mode. 
In agreement with above discussion of the upward moving light
selection criteria (\ref{eq1}) we found for nearly all laser events 
$t_{\mbox{\footnotesize min}}>-$10 ns.

The high transparency and the low scattering of Baikal water
allow the reconstruction of coordinates and intensity
of bright light flashes.
The laser experiment allowed to verify the reconstruction procedure
of shower position and intensity. We found a vertical laser
coordinate precision of $\sim$1 m and relative intensity
precision of $\sim$10 \% \cite{SHAI}.

\subsection{High-energy neutrino simulation}
The number of expected events $N_{\nu_i}$ during
observation time T is: 
$$
N_{\nu_i}=T \int d\vec{\Omega}\int dE_{h}V_{eff}
(\vec{\Omega},E_{sh})\times
$$
\begin{equation}
\label{h4}
\sum_k \int
 N_{A} \rho_{H_2O} \frac{d\sigma_{\nu k}}{dE_{h}}
\Phi_{\nu_i}(\vec{\Omega},E_{\nu},X)dE_{\nu}
\end{equation}
$$
X(\vec{\Omega})=\int_{0}^{L} \rho_{earth}(l)dl, 
$$
where $\Phi_{\nu_i}(\vec{\Omega},E_{\nu},X)$ is the flux of high energy 
neutrinos with energy $E_{\nu}$ in the vicinity of the detector,
$\vec{\Omega}$ - the neutrino direction, $X(\vec{\Omega})$ - the thickness 
of matter encountered by the neutrino on its passage through the Earth,
$E_{\mbox{\footnotesize h}}$ - the energy transferred to the hadron, 
$E_{\mbox{\footnotesize sh}}$ - total energy of secondary showers, 
$V_{\mbox{\footnotesize eff}}(\vec{\Omega},E_{\mbox{\footnotesize sh}})$ - 
the detection volume. 
The index $\nu_i$ indicates the neutrino type
($\nu_i=\nu_e, \bar{\nu_e}, \nu_{\mu},\bar{\nu_{\mu}},
\nu_{\tau},\bar{\nu_{\tau}}$) and
$k=$1,2 corresponds to CC- and NC-interactions respectively.
$N_{A}$ is the Avogadro number and $\rho_{H_2O}$ the water density. 

For $\nu_e$ and $\nu_{\mu}$,  the flux $\Phi_{\nu}$ 
satisfies the following transport equation:
$$
\frac{d\Phi_{\nu}(E_{\nu})}{dX}=-N_A \sigma_{tot}\Phi_{\nu}(E_{\nu})+
$$
\begin{equation}
\label{h04}
N_A\int_{E_{\nu}}^{\propto}dE'_{\nu}
\Phi_{\nu}(E'_{\nu})\frac{d\sigma^{NC}}{dE_{\nu}}(E'_{\nu},E_{\nu})
\end{equation}
where $\sigma_{tot}=\sigma_{CC}+\sigma_{NC}$.

For $\nu_{\tau}$,  the tau neutrino and tau lepton fluxes 
satisfy the equations:
$$
\frac{d\Phi_{\nu_{\tau}}(E_{\nu_{\tau}})}{dX}=
-N_A\sigma_{tot} \Phi_{\nu_{\tau}}(E_{\nu_{\tau}})+
$$
$$
N_A\int_{E_{\nu_{\tau}}}^{\propto}dE'_{\nu_{\tau}}
\Phi_{\nu_{\tau}}(E'_{\nu_{\tau}})
\frac{d\sigma^{NC}}{dE_{\nu_{\tau}}}(E'_{\nu_{\tau}},E_{\nu_{\tau}})
$$
\begin{equation}
\label{h05}
+\int_{E_{\nu_{\tau}}}^{\propto}dE_{\tau}
\frac{\Phi_{\tau}(E_{\tau})}{\lambda^{dec}_{\tau}}\frac{dn}{dE_{\nu_{\tau}}}(E_{\tau},E_{\nu_{\tau}})
\end{equation}
$$
\frac{d\Phi_{\tau}(E_{\tau})}{dX}=-\frac{\Phi_{\tau}(E_{\tau})}
{\lambda^{dec}_{\tau}(E_{\tau})}+
$$
$$
N_A\int_{E_{\tau}}^{\propto}dE_{\nu_{\tau}}
\Phi(E_{\nu_{\tau}})\frac{d\sigma^{CC}}{dE_{\tau}}(E_{\nu_{\tau}},E_{\tau})
$$
where $\lambda^{dec}_{\tau}=\gamma c \tau_{\tau} \rho$ is the decay length 
of the $\tau$-lepton, $dn/dE_{\nu_{\tau}}$ - $\nu_{\tau}$ production 
probability. The transport equation for
the $\tau$-lepton is valid under the assumption that
$\lambda^{dec}_{\tau}/\lambda_{\tau}<<1$, where $\lambda_{\tau}$ -
the interaction length of $\tau$ propagation through the medium. 
This assumption is valid for energies $E_{\tau}<$10$^8$ GeV.

\begin{figure*}
\includegraphics*[width=.5\textwidth,height=6.0cm]{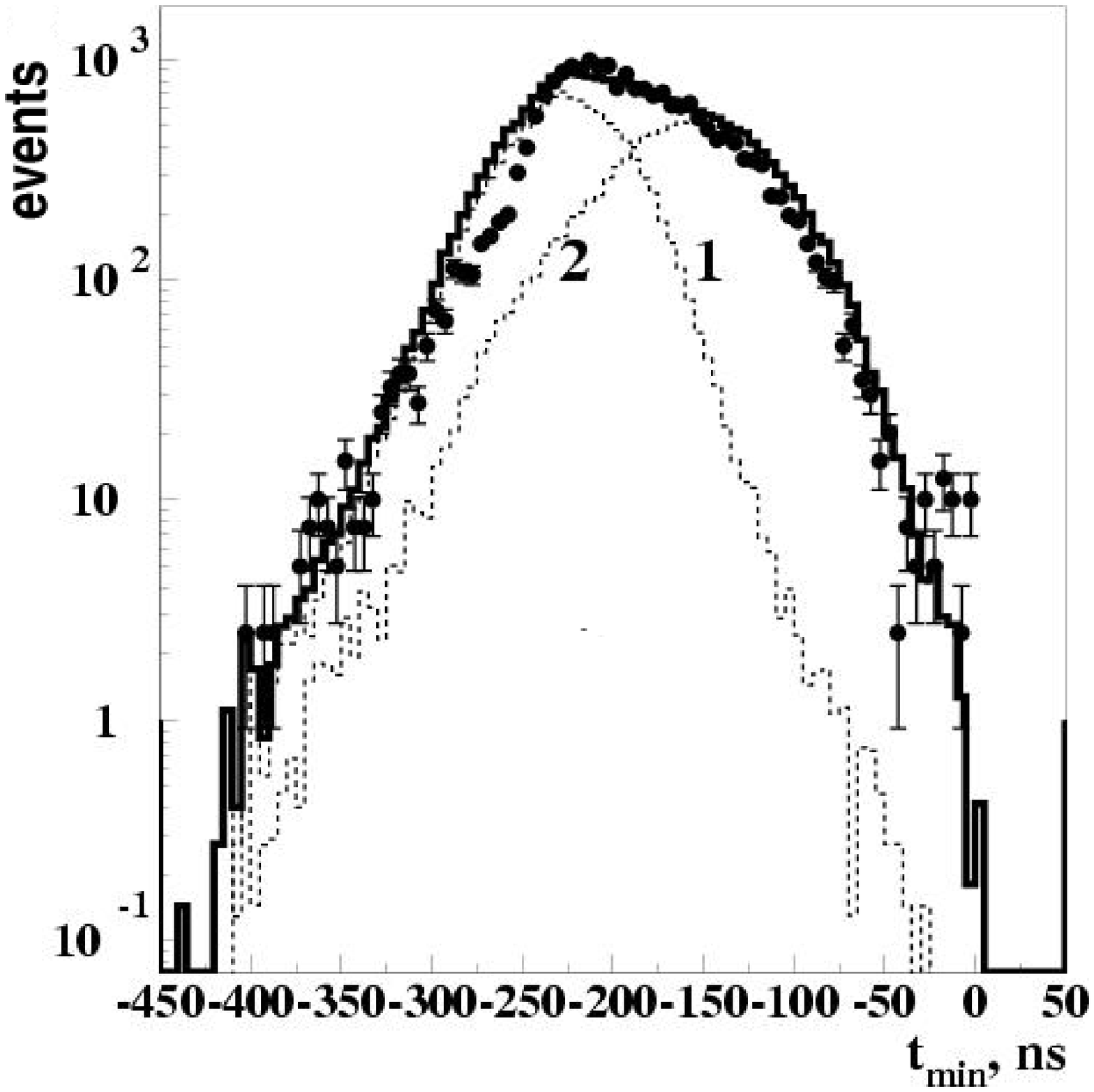}
\hfill
\includegraphics*[width=.5\textwidth,height=6.0cm]{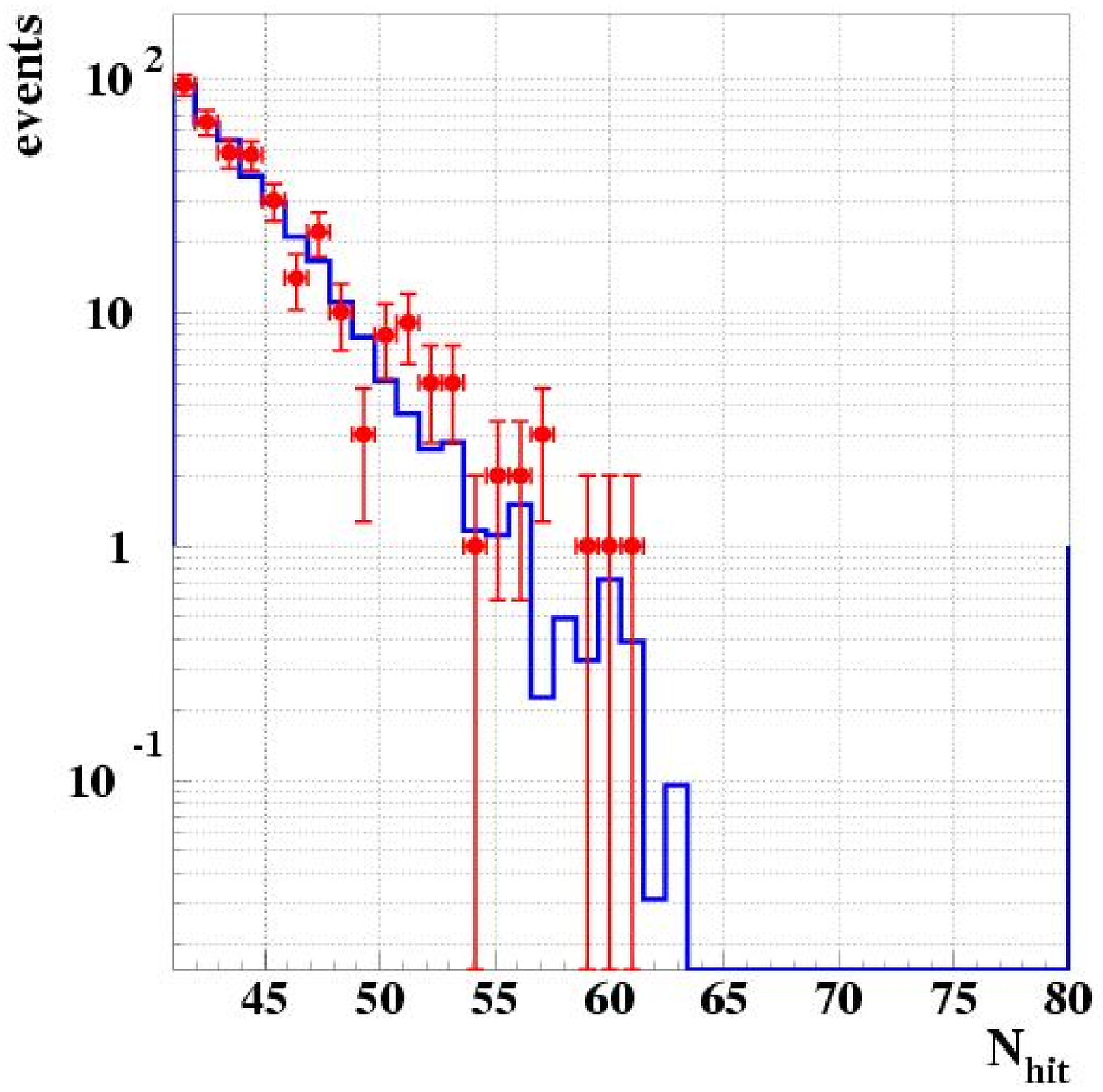}
\\
\caption{Left panel: 
the $t_{\mbox{\footnotesize min}}$ distributions for 
experiment (dots) and atmospheric
muon simulation (bold histogram) for $N_{\mbox{\small hit}}>$40. 
Histograms 1 and 2 correspond to atmospheric muon simulation
with $\cos \theta >0.8$ and $\cos \theta <0.8$, respectively.
Right panel: the $N_{\mbox{\small hit}}$ distributions for 
experiment (dots) and atmospheric
muon simulation (histogram) with $t_{\mbox{\footnotesize min}}>$-100 ns.}
\label{fig9}
\end{figure*}
A MC-code is used to solve equations (\ref{h04}) - (\ref{h05}),
with the boundary conditions for neutrino fluxes 
$\Phi_{\nu_i}(E,0)=A_{\nu_i}f_{\nu_i}(E)$,
where $f_{\nu_i}(E)$ is a diffuse AGN-like flux or other 
predicted UHE neutrino fluxes, and a $A_{\nu_i}$ a normalization 
coefficient. For tau leptons $\Phi_{\tau}(E,0)=0$.
For neutrino interactions and tau-neutrino regeneration we used
cross-sections from \cite{Reno96,Lip93}.
The neutrinos are propagated through the Earth assuming
the density profile of the Preliminary Reference Earth Model \cite{EARTH}. 
Although a flavor ratio of $\nu_e:\nu_{\mu}:\nu_{\tau}\approx$1:2:0
is predicted for generic neutrino fluxes at cosmic sources,
equal fractions of all three neutrino flavors are expected
at Earth because of neutrino oscillations.
Throughout this paper we assumed 
a neutrino flavor ratio at Earth of $\nu_e:\nu_{\mu}:\nu_{\tau}=$1:1:1
and the same shape  of energy spectra $f(E_{\nu})$ for all neutrino flavors,
as well as a flux ratio for neutrino and antinutrino of 
$\nu/\bar{\nu}=$1\footnote{A violation of this assumption (e.g. for neutrino production
in $p\gamma$ interactions) has a small influence on the result due to the similarity
of $\nu$ and $\bar{\nu}$ cross-sections in our energy range.}.

The detector response to Cherenkov radiation of high energy
cascades was simulated taking into account the effects
of absorption and scattering of light 
as well as light velocity dispersion in water \cite{KUZ,DZH1,DZH2}.
We also implemented the longitudinal development of cascades.
For electron cascades with $E_{\mbox{\footnotesize sh}}>$2$\times$10$^7$ GeV
and for hadronic cascades with $E_{\mbox{\footnotesize sh}}>$10$^9$ GeV, 
the increase
in cascade length due to the LPM effect \cite{LPM1}
was approximated as $E^{1/3}$ according to Ref. \cite{LPM}.

\subsection{Atmospheric muon simulation}
Downward going atmospheric muons 
are the most important source of background. 
The simulation chain of these muons starts with 
cosmic ray air shower generation using the CORSIKA
program \cite{CORSIKA} with the QGSJET \cite{QGSJET}
interaction model and
the primary composition and spectral slopes
for individual elements taken from \cite{Smooth}.
Atmospheric muons are propagated through the water
using the MUM program \cite{MUM}. 
During passage through the detection volume the detector response 
to Cherenkov light from all muon energy loss processes
is simulated.

For illustration of the consistency of experimental data
with simulation we present in fig. \ref{fig9} (left panel) 
the $t_{\mbox{\footnotesize min}}$ distributions for
simulated background events (histograms), as well as 
for experimental events with a hit multiplicity 
$N_{\mbox{\small hit}}>40$ taken during 41.8 live days in 1999. Curves 1 and 2  
correspond to atmospheric muons with zenith angles 
$\cos \theta>0.8$ and $\cos \theta<0.8$, respectively. 
We find good agreement for all $t_{\mbox{\footnotesize min}}$ values, with a small 
deficit only for vertical downward going muons in the interval -300, -230 ns
(i.e. far-off the cut value -10 ns). This deficit is due to the difficulty 
to precisely determine the low OM detection efficiency of downward looking OMs 
for straight backward light illumination by vertical atmospheric muons.
Figure \ref{fig9} (right panel) shows the $N_{\mbox{\small hit}}$ distribution 
of events with $t_{\mbox{\footnotesize min}}>$-100 ns as 
well as background simulation (histogram). 
We conclude that experimental $t_{\mbox{\footnotesize min}}$ 
and $N_{\mbox{\small hit}}$ 
distributions are consistent with distributions expected for atmospheric muons.

\begin{figure*}
\includegraphics*[width=.5\textwidth,height=6.0cm]{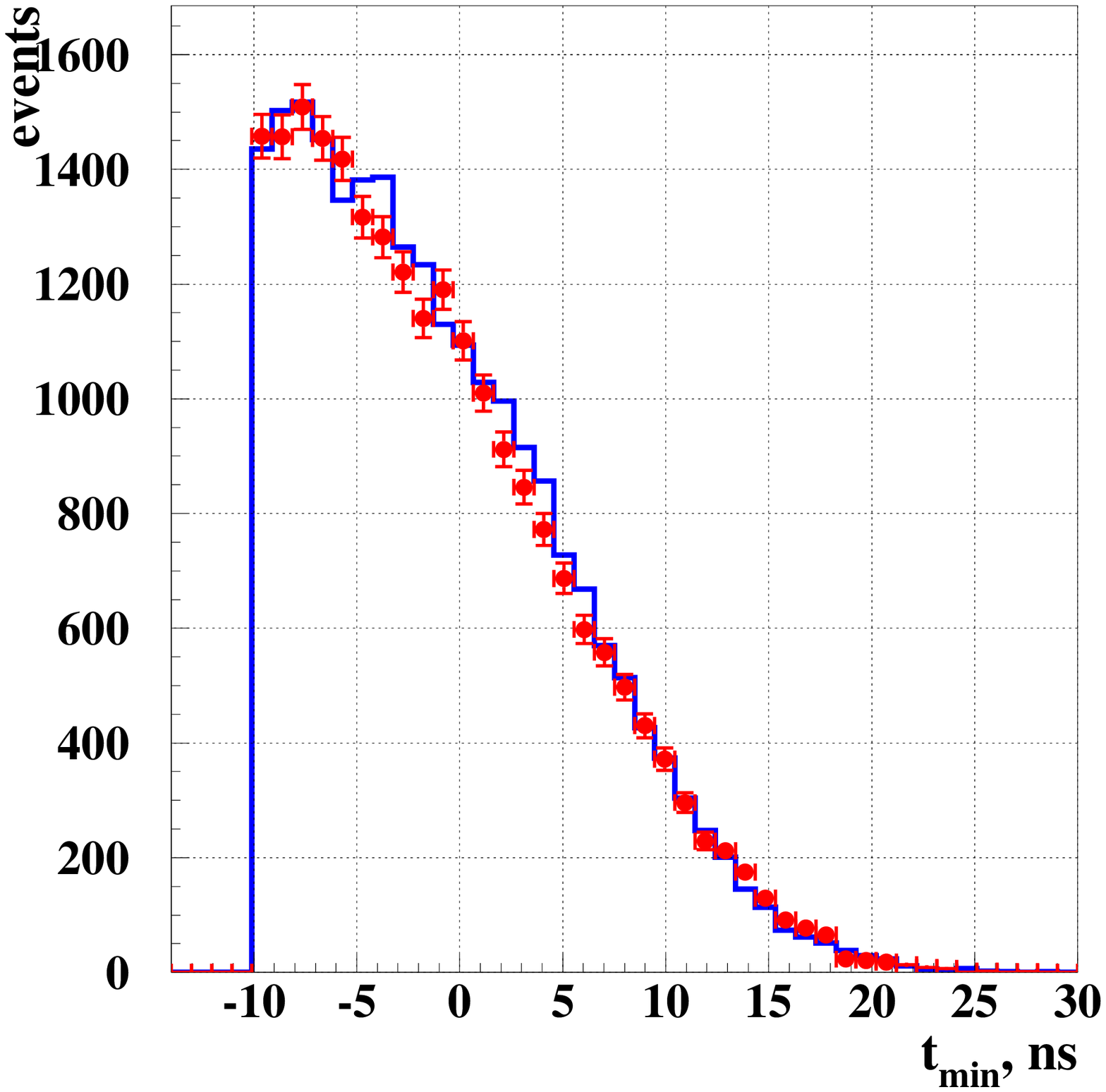}
\hfill
\includegraphics*[width=.5\textwidth,height=5.4cm]{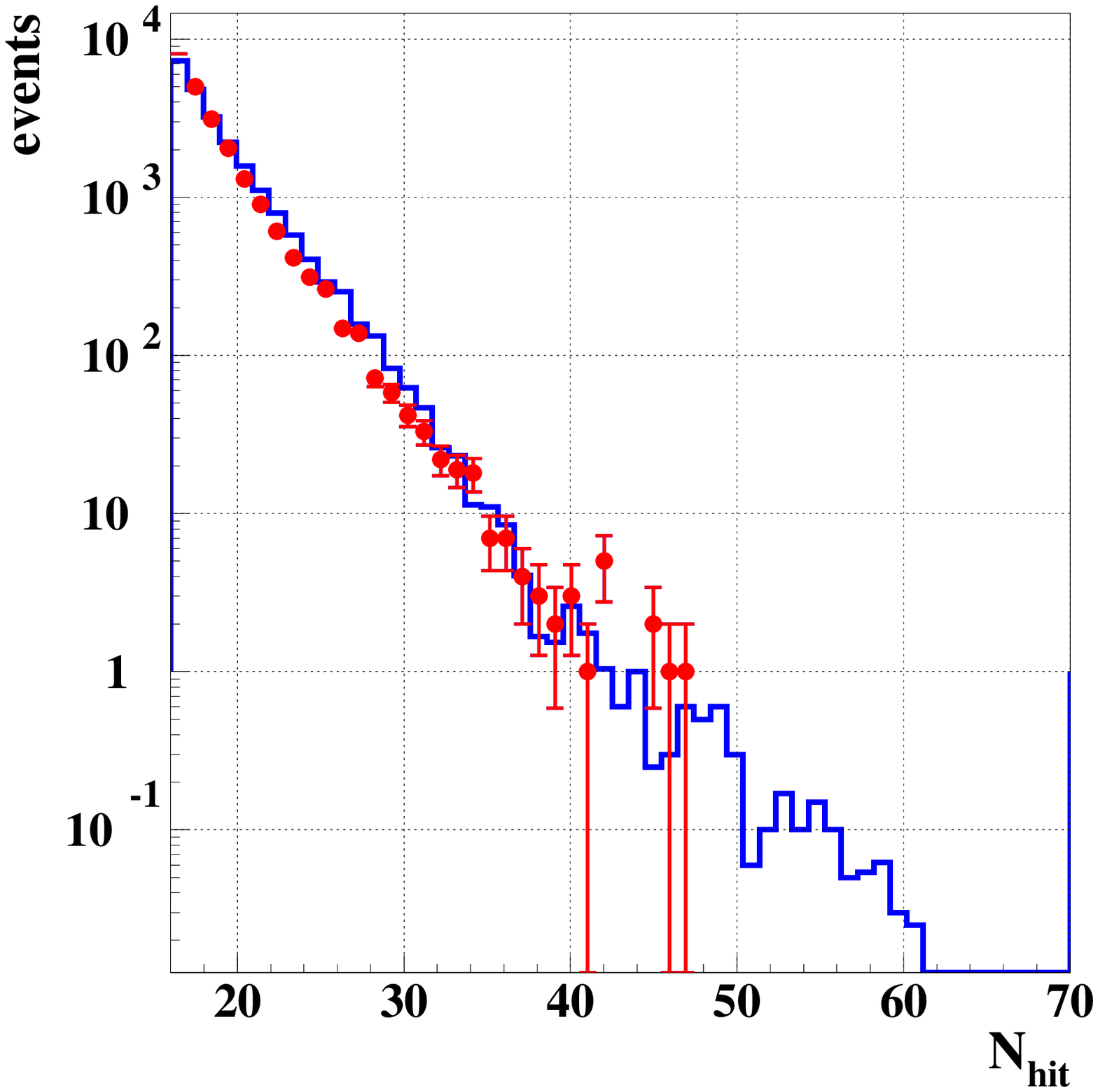}
\\
\caption{Left panel: 
the $t_{\mbox{\footnotesize min}}$ distribution of experimental events (dots)
which survive cut (\ref{eq1}) as well as the expected 
distribution of simulated background events (histogram). Right panel:
the $N_{\mbox{\small hit}}$ distribution of experimental events (dots)
as well as the background prediction (histogram) after cut (\ref{eq1}).}
\label{fig11}
\end{figure*}

\section{Data selection and analysis}

Within the 1038 days of the detector live time
between April 1998 and February 2003, 
$3.45 \times 10^8$ events with $N_{\mbox{\small hit}} \ge 4$ have been recorded. 
For this analysis we used 22597 events with hit channel multiplicity
$N_{\mbox{\small hit}}>$15 which  obey the condition (\ref{eq1}). 
For these events upward looking channels as well as channels
which are operated in the 1-PM/channel mode (see section 2) have 
been excluded from the following analysis.

\begin{table}[htb]
\caption{{\it NT200} effective configurations.}
\label{tab1}
  \begin{tabular}{@{}ccccc}
\hline
Conf. & $\bar{N}_{\mbox{\footnotesize op}}$ & $T$ &
$N_{\mbox{\footnotesize ev}}$ & 
$N_{\mbox{\footnotesize hit}}^{\mbox{\footnotesize max}}$ \\
  {} &{} &{(days)} &{} &{} \\
  \hline
  1 & 71 & 316 & 12146 & 47 \\
  2 & 59 & 612 & 9473 & 42 \\
  3 & 46 & 110 & 978 & 32 \\
  \hline
\end{tabular} 
\end{table}

During the 1038 days {\it NT200} 
took data in various configurations, which have evolved
due to  groups of OMs failing.
Neglecting few-OM differences, the data can be grouped according to three 
basic configurations. The average number of working channels 
$\bar{N}_{\mbox{op}}$, the data taking time $T$, the number of detected events 
$N_{\mbox{ev}}$, as well as the largest multiplicity of hit channels for 
detected events $N_{\mbox{\small hit}}^{\mbox{\small max}}$ 
are shown in Table \ref{tab1}.

\begin{figure*}
\includegraphics*[width=.5\textwidth,height=8.0cm]{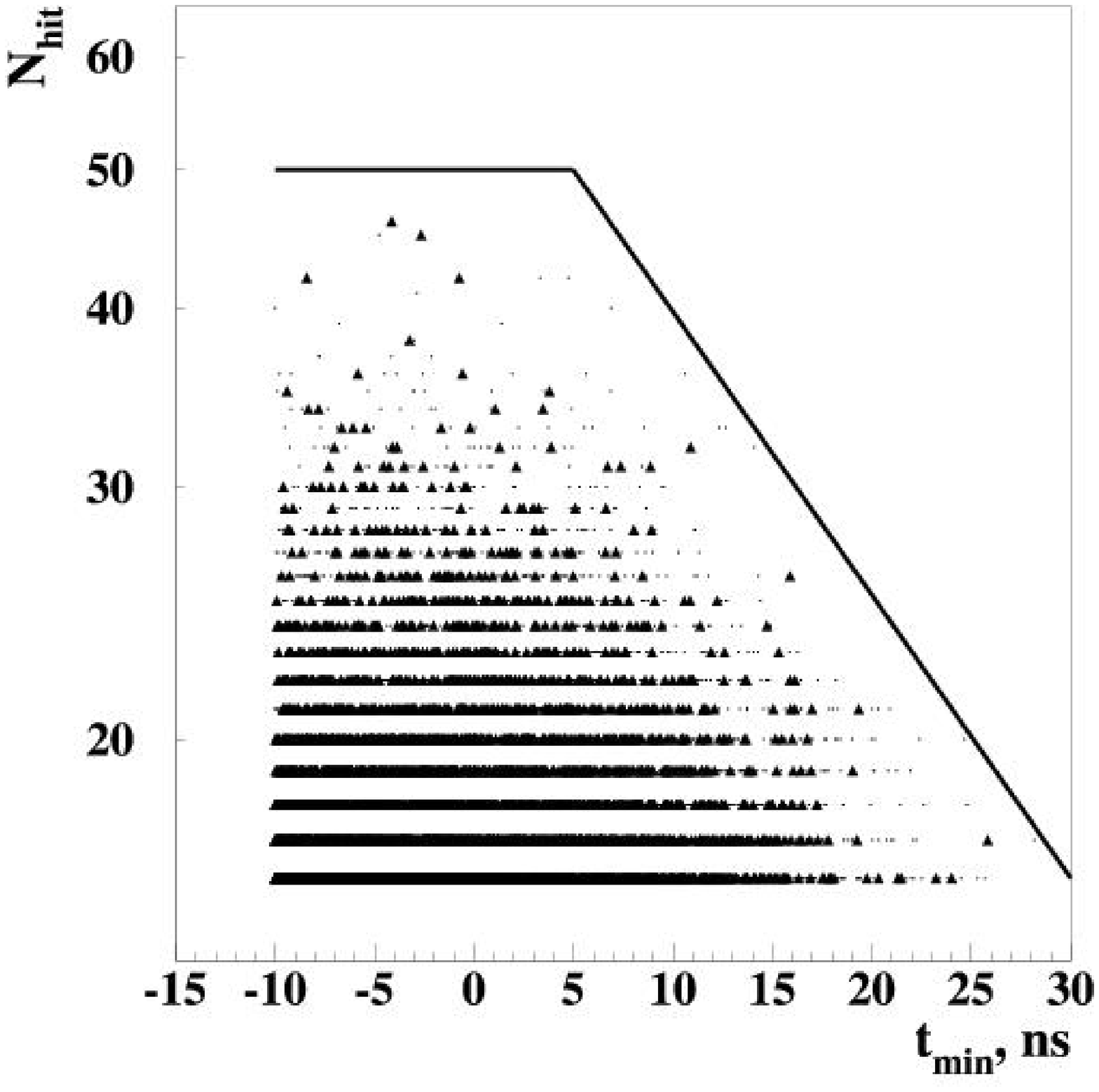}
\hfill
\includegraphics*[width=.5\textwidth,height=8.0cm]{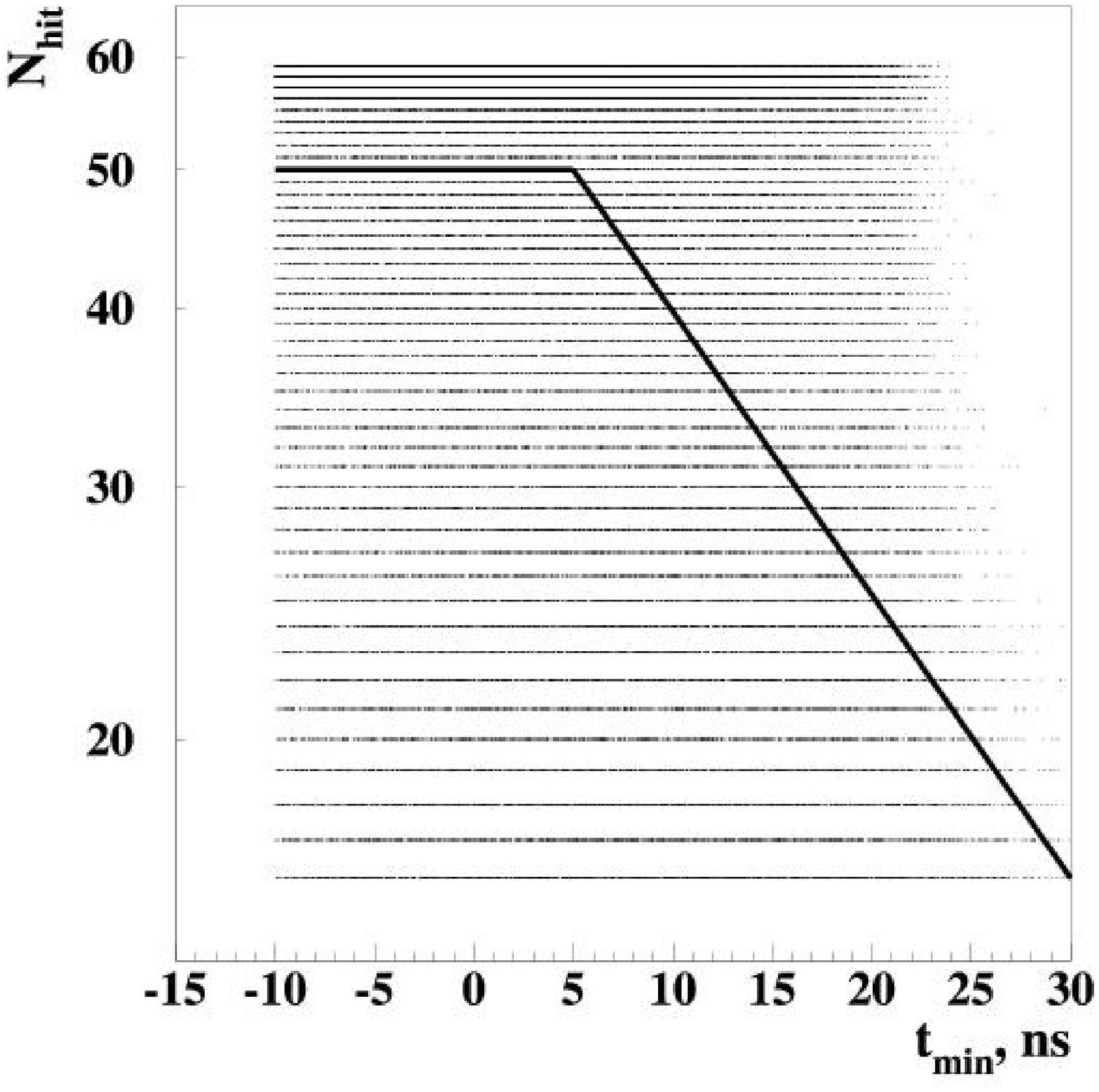}
\\
\vspace*{-0.3cm}
\caption{Left panel: distributions of experimental (triangles) and expected 
background (dots) events in the ($t_{\mbox{\footnotesize min}},N_{\mbox{\small hit}}$)-plane.
Right panel: same distribution for events expected from high energy 
cascades. Results are shown for configuration 1 (Table \ref{tab1}). The line
shows cut condition (see Table \ref{tab2}).}
\label{fig12}
\end{figure*}
\begin{table*}[htb]
\caption{Cut conditions in the ($t{\mbox{\footnotesize min}}, N_{\mbox{\small hit}}$)-plane
used for the selection of neutrino induced events ($t{\mbox{\footnotesize min}}$ in ns).}
\label{tab2}
\renewcommand{\tabcolsep}{1.5pc} 
\renewcommand{\arraystretch}{1.2} 
  \begin{tabular}{@{}llll}
\hline
Conf. 1 & -10$<t_{\mbox{\footnotesize min}} \leq$5 & 5 $<t_{\mbox{\footnotesize min}} \leq$30 & $t_{\mbox{\footnotesize min}} >$30 \\
 &$N_{\mbox{\small hit}} \geq$ 50 &$N_{\mbox{\small hit}} \geq$ 10$^{(1.8-0.02t_{min})}$ &$N_{\mbox{\small hit}} \geq$ 16 \\
Conf. 2 & -10 $<t_{\mbox{\footnotesize min}} \leq$10 & 10$<t_{\mbox{\footnotesize min}} \leq$30 & $t_{\mbox{\footnotesize min}} >$30 \\
 &$N_{\mbox{\small hit}} \geq$  44 &$N_{\mbox{\small hit}} \geq$ 10$^{(1.865-0.0223t_{min})}$ &$N_{\mbox{\small hit}} \geq$ 16 \\
Conf. 3 & -10 $<t_{\mbox{\footnotesize min}} \leq$10 & 10 $<t_{\mbox{\footnotesize min}} \leq$30 & $t_{\mbox{\footnotesize min}} >$30 \\
 &$N_{\mbox{\small hit}} \geq$  33 &$N_{\mbox{\small hit}} \geq$ 42-0.9$t_{\mbox{\footnotesize min}}$ &$N_{\mbox{\small hit}} \geq$ 16 \\
\hline
\end{tabular} 
\end{table*}

Figure \ref{fig11} shows the $t_{\mbox{\footnotesize min}}$ and $N_{\mbox{\small hit}}$ 
distributions for experiment (dots) and background simulation (histograms).
We conclude that after application cut (\ref{eq1}) data are consistent with simulated background
for both $t_{\mbox{\footnotesize min}}$ and $N_{\mbox{\small hit}}$ distributions.
No statistically significant excess above the background 
from atmospheric muons has been observed. 

To maximize the sensitivity to a neutrino signal we introduce a cut in the 
($t_{\mbox{\footnotesize min}},N_{\mbox{\small hit}}$) phase space.
Figure \ref{fig12} (left panel) shows
the population of the ($t_{\mbox{\footnotesize min}},N_{\mbox{\small hit}}$) phase space for
experimental events (triangles) as well as for background simulation (dots).
The distribution  for neutrino induced 
events (dots) is shown in the right panel of fig. \ref{fig12} (for spectral index $\gamma=-$2).
We note that background events populate the lower-left part of  the plot, in contrast to signal
events.
The bounds which fence the ($t_{\mbox{\footnotesize min}}, 
N_{\mbox{\small hit}}$) area which we assigned to
background are listed in Table \ref{tab2} for the three {\it NT200} 
configurations. 

With no experimental events outside the area populated by background 
events in the ($t_{\mbox{\footnotesize min}}, 
N_{\mbox{\small hit}}$) phase space
we can derive upper limits on the fluxes of high energy neutrinos as
predicted by different models of neutrino sources. 

\begin{figure}
\begin{center}
\includegraphics*[width=.5\textwidth,height=7.0cm]{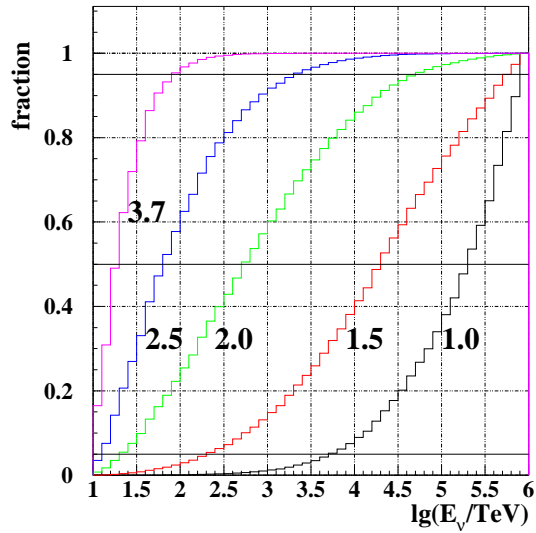}
\end{center}
\caption{The fraction of expected events induced by diffuse
$\nu_e$ fluxes with spectral indices $\gamma$=1, 1.5, 2, 2.5 and 3.7
within energy range 10 TeV$<E<E_{\nu}$.
} 
\label{fig13}
\end{figure}

The neutrino detection energy range of {\it NT200} which contains, 
for instance, 90\% of expected events, depends on the energy shape of 
the neutrino flux. Figure \ref{fig13} shows the fractions of
expected events induced by diffuse neutrino fluxes following an 
$E^{-\gamma}$ shape with spectral indices $\gamma$=1, 1.5, 2, 2.5 and 3.7, 
for energies ranges 10 TeV$<E<E_{\nu}$. The effective detection range is 
shifting towards higher energies with  decreasing $\gamma$. Table \ref{tab3} 
shows the energy ranges for effective detection of neutrino fluxes with 
different $\gamma$, as well as  the median values of the energy 
distributions.     

\begin{table}[htb]
\caption{Detection energy range and median energy for different spectra.}
\label{tab3}
  \begin{tabular}{@{}ccc}
\hline
$\gamma$ & $\Delta E$, TeV & $E_{\mbox{med}}$, TeV  \\
  \hline
  1.0 & 5$\times$10$^3$ - 7$\times$10$^5$ & 2.2$\times$10$^5$  \\
  1.5 & 2$\times$10$^2$ - 5.6$\times$10$^5$ & 2.0$\times$10$^4$  \\
  2.0 & 22 - 5.0$\times$10$^4$ & 5.6$\times$10$^2$  \\
  2.5 & 14 - 2.0$\times$10$^3$ & 63  \\
  3.7 & 10 - 89 & 20  \\
  \hline
\end{tabular} 
\end{table}

With increasing energy, neutrinos are stronger absorbed when
propagating through the Earth, and upward going fluxes are
suppressed. Our search strategy accepts neutrinos from all directions.
With increasing neutrino energy the angular distributions are shifted
towards smaller zenith angles. Zenith angle ranges $\Delta \theta$ 
which contain 90\% of expected events induced by an $E^{-2}$ neutrino flux
are 75$^o$ - 150$^o$ and 40$^o$ - 90$^o$ for energy ranges 10 TeV - 100 TeV
and 10$^5$ - 10$^6$ TeV, respectively.

\begin{figure*}
\includegraphics*[width=.45\textwidth,height=6.3cm]{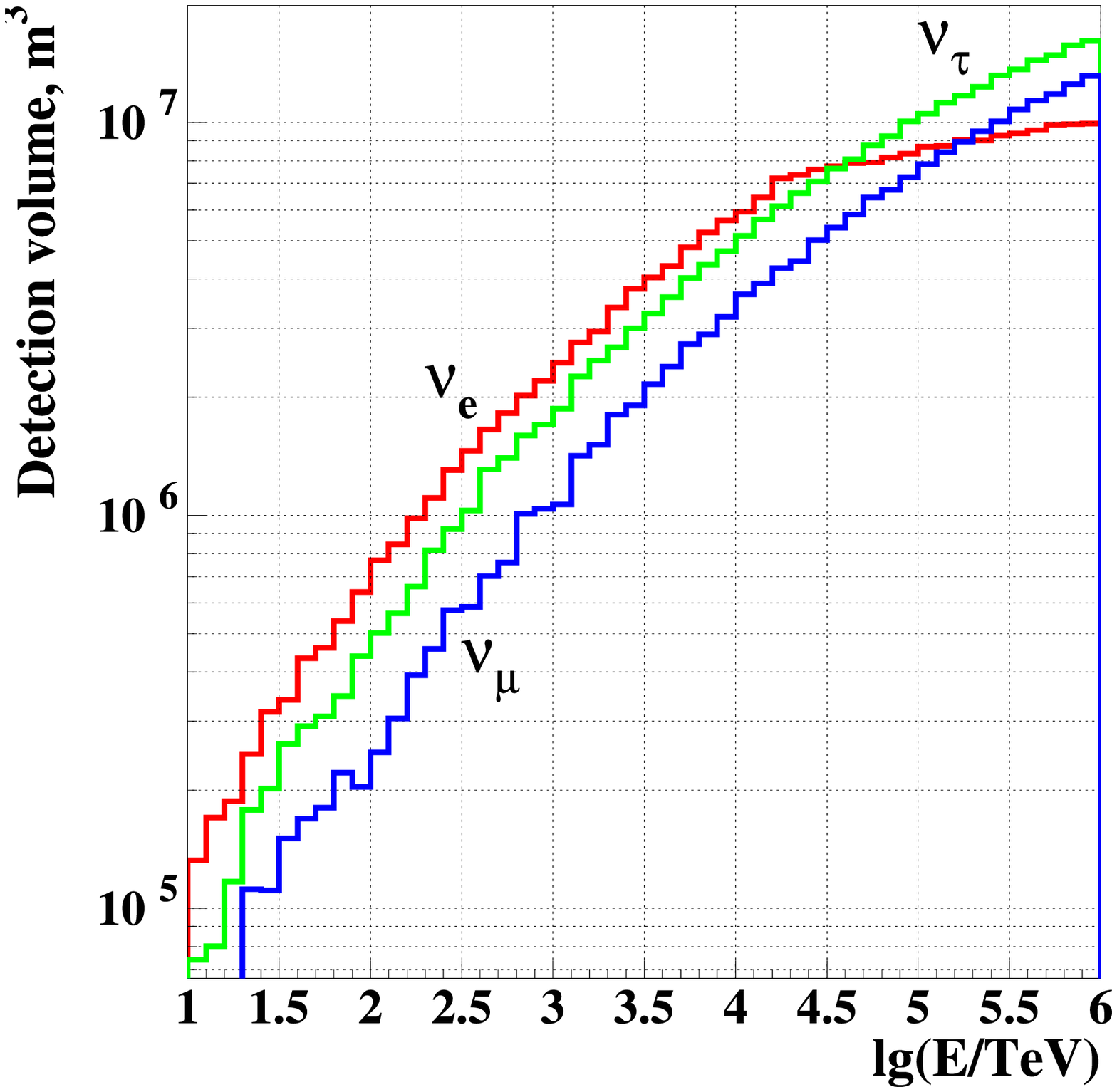}
\hfill
\includegraphics*[width=.55\textwidth,height=7.0cm]{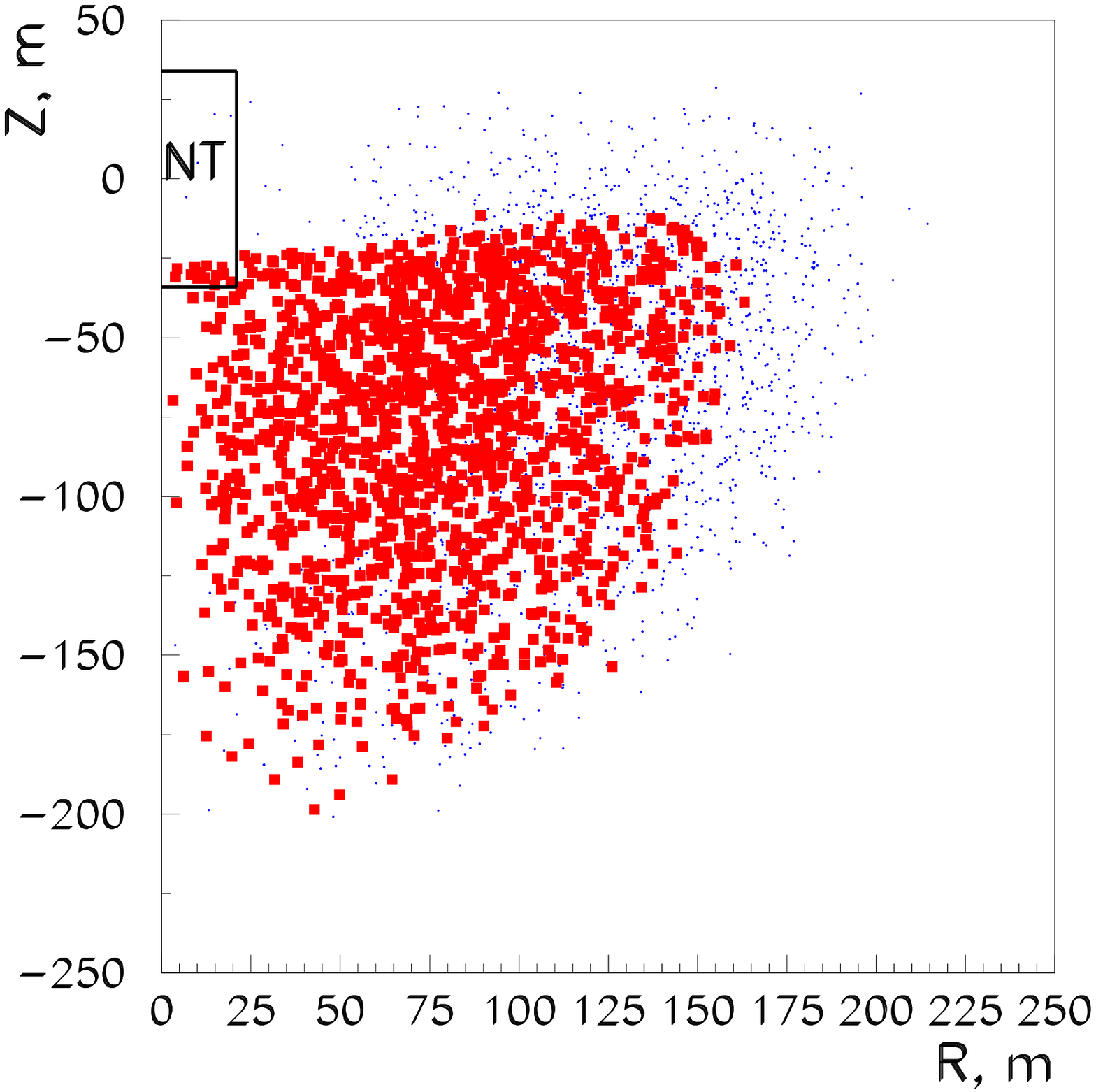}
\\
\caption{Left panel: 
energy dependence of detection volumes. 
Right panel: 
coordinates of $\nu_e$ interaction vertices
for events which fulfill cut (\ref{eq1}) (dots) and all 
final cuts for neutrino event selection (rectangles), respectively.
The ordinate gives the vertical position, the abscissa the radial
distance, with (z=0, R=0) corresponding to the center of {\it NT200}.
The small rectangle marks the geometrical volume of {\it NT200}.
}
\label{fig15}
\end{figure*}
The detection volumes V$_{\mbox{\footnotesize eff}}$ for all three neutrino flavors after 
all cuts were calculated as a function of neutrino energy and zenith angle 
$\theta$. The energy dependence of the detection volumes,  
averaged over all neutrino arrival directions, are shown in fig. \ref{fig15} (left panel). 
The value of $V_{\mbox{\footnotesize eff}}$ rises from $\sim$10$^5$ m$^3$ for 10 TeV up to 
(4-6)$\times$10$^6$ m$^3$ for $10^4$ TeV and significantly exceeds the 
geometrical volume $V_{\mbox{\footnotesize g}}\approx$ 10$^5$ m$^3$ of {\it NT200}. This is
due to the low light scattering and the nearly not distorted light fronts
from Cherenkov waves originating far outside the geometrical
volume. In the case of $\nu_e$ detection, the  volume saturates 
for $E_{\nu_e}>$10$^4$ TeV because of the LPM effect.
Figure \ref{fig15} (right panel)  illustrates the difference between $V_{\mbox{\footnotesize eff}}$ and 
$V_{\mbox{\footnotesize g}}$. 
Shown here are the coordinates of neutrino interaction vertices 
for events which survive cuts (\ref{eq1}) (dots) and 
all cuts (rectangles), assuming an $E^{-2}$ spectrum. 

Systematic uncertainties in the optical properties of  water, the absolute 
detector sensitivity and the neutrino cross sections at high energies
influence the number of expected signal events. The uncertainty of 10\% in absorption 
length results in 20\% uncertainty in the number of expected events. 
The uncertainty of 14\% in the sensitivity of OMs results in 10\% uncertainty 
in the number of expected events. Below 10$^{16}$ eV, all standard sets of 
parton distributions yield very similar cross sections. Above this energy, 
the cross sections are sensitive to assumptions made about the behavior for
$x \rightarrow 0$. The uncertainties of cross sections are less than 
10\% below 10$^{18}$ eV \cite{Reno96,Gluck,Martin} and result in uncertainties 
of 8\% in the number of expected events. 
Treating these errors as independent and adding them
quadratically, the signal uncertainty becomes 24\%.

\vspace{-0.5cm}
\section{Limits on the high energy neutrino fluxes}
\vspace{-0.5cm}
\begin{table*}[htb]
\caption{Expected number of events $N_{\mbox{\footnotesize model}}$ and model rejection factors for models of
astrophysical neutrino sources. The assumed upper limit on the
number of signal events with all uncertainties incorporated is $n_{90\%}=2.5$}
\label{tab5}
  \begin{tabular}{@{}lccccc|c}
\hline
 & \multicolumn{5}{c|}{BAIKAL} & AMANDA \cite{AMANDAHE,AMANDAMU}\\
\hline
Model & $\nu_e$ & $\nu_{\mu}$ & $\nu_{\tau}$ & $\nu_e+\nu_{\mu}+\nu_{\tau}$ & $n_{90\%}/N_{\mbox{\footnotesize model}}$ & $n_{90\%}/N_{\mbox{\footnotesize model}}$  \\
\hline
  10$^{-6}\times E^{-2}$ & 1.33 & 0.63 & 1.12 & 3.08 & 0.81 & 0.86  \\
  SS Quasar \cite{SS} & 4.16 & 2.13 & 3.71 & 10.00 & 0.25 & 0.21  \\
  SP u  \cite{SP}& 17.93 & 7.82 & 14.43 & 40.18 & 0.062 & 0.054  \\
  SP l \cite{SP}& 3.14 & 1.24 & 2.37 & 6.75 & 0.37 & 0.28  \\
  P $p\gamma$ \cite{P}& 0.81 & 0.53 & 0.85 & 2.19 & 1.14 & 1.99  \\
  M $pp+p\gamma$ \cite{M} & 0.29 & 0.22 & 0.35 & 0.86 & 2.86 & 1.19  \\
  MPR \cite{MPR}& 0.25 & 0.14 & 0.24 & 0.63 & 4.0 & 4.41  \\
  SeSi \cite{SeSi} & 0.47 & 0.26 & 0.44 & 1.18 & 2.12 & -  \\
  \hline
\end{tabular} 
\end{table*}
Since no events have been observed which pass the final cuts
(see Table \ref{tab2}), upper limits on the diffuse flux of extraterrestrial 
neutrinos are calculated. For a 90\% confidence level an upper limit 
on the number of signal events of $n_{90\%}=$2.5  is obtained according to Conrad et al. 
\cite{CONRAD} with the unified Feldman-Cousins ordering \cite{FC}. We assume 
an uncertainty in signal detection of 24\% and a background of zero events 
(which leads to a conservative estimation of $n_{90\%}$ according to 
the Feldman-Cousins approach).

The expected number of signal events $N_{\nu_i}$ for
any assumed flux $\Phi_{\nu_i}(E)$  of neutrinos of flavor $i$,
is given by expression (\ref{h4}).
A model of astrophysical neutrino sources, for which the total number
of expected events, $N_{\mbox{\footnotesize model}}$, is large than 
$n_{90\%}$, is ruled out at 90\% CL. $N_{\mbox{\footnotesize model}}$ is
given as $N_{\mbox{\footnotesize model}}=\sum N_{\nu_i}$.
Table \ref{tab5} represents event rates and model rejection factors (MRF) 
$n_{90\%}/N_{\mbox{\footnotesize model}}$ 
for models of astrophysical neutrino sources 
obtained from our search. Recently, similar results have been 
presented by the AMANDA collaboration \cite{AMANDAHE,AMANDAMU},
model rejection factors are shown in Table \ref{tab5}.

The models by Stecker and Salamon \cite{SS} labeled ``SS Q'',
as well as the models by Szabo and Protheroe \cite{SP} ``SP u''
and ``SP l'' represent models for neutrino production in the central
region of Active Galactic Nuclei. As can be seen from Table \ref{tab5},
these models are ruled out with $n_{90\%}/N_{\mbox{\footnotesize model}} 
\approx$ 0.06 - 0.4.
Further shown are models for neutrino production in AGN jets:
calculations by Protheroe \cite{P} and by Mannheim \cite{M},
which include neutrino production through $pp$ and $p\gamma$ collisions
(models ``P $p\gamma$'' and ``M $pp+p\gamma$'', respectively),
as well as an evaluation of the maximum flux due to a superposition of
possible extragalactic sources by Mannheim, Protheroe and Rachen
\cite{MPR} (model ``MPR'') and a prediction for the diffuse
flux from blazars by Semikoz and Sigl \cite{SeSi} ``SeSi''.
The latter models for blazars are currently not excluded.

\begin{figure*}
\includegraphics*[width=.45\textwidth,height=7.0cm]{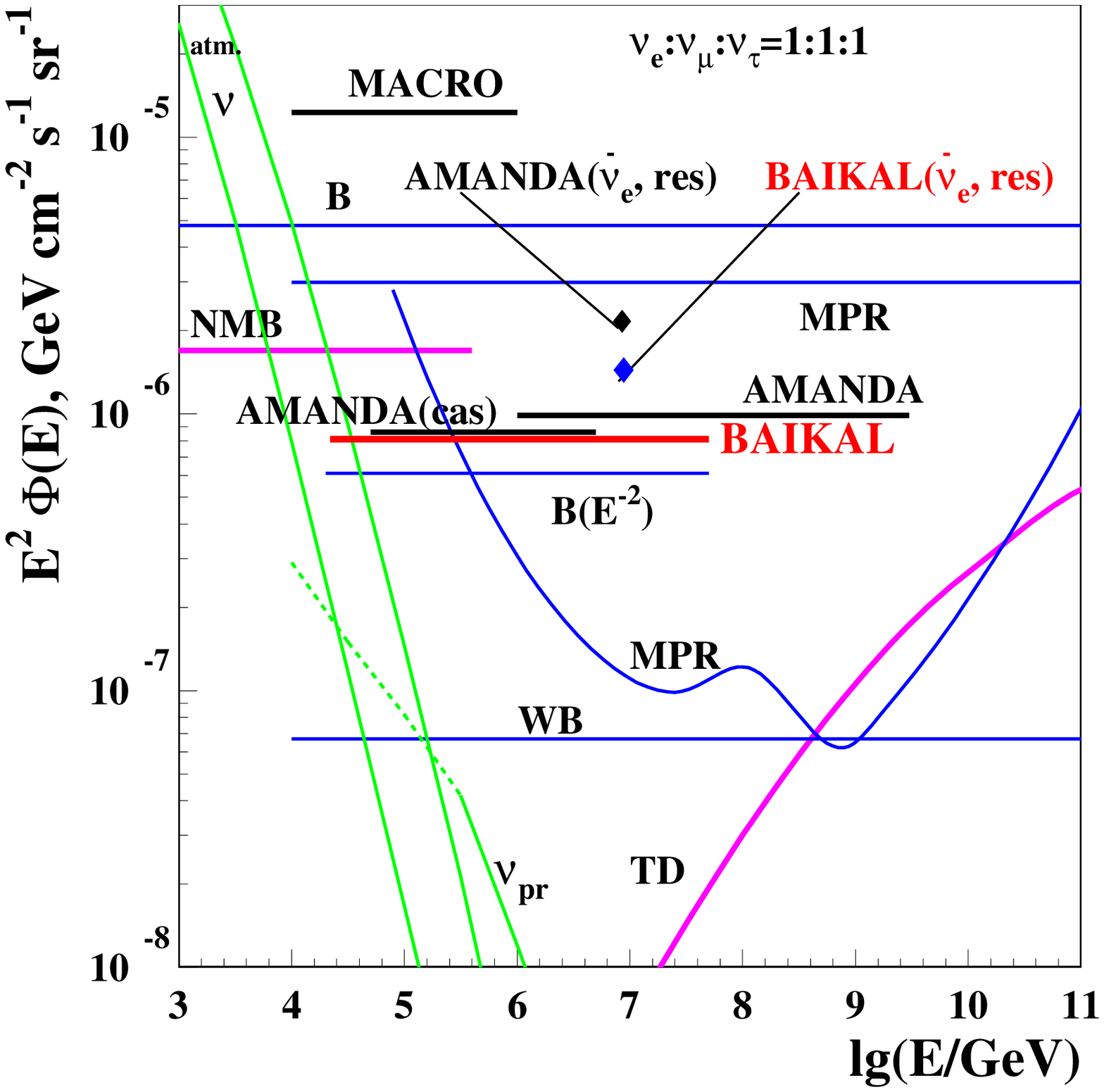}
\hfill
\includegraphics*[width=.45\textwidth,height=7.0cm]{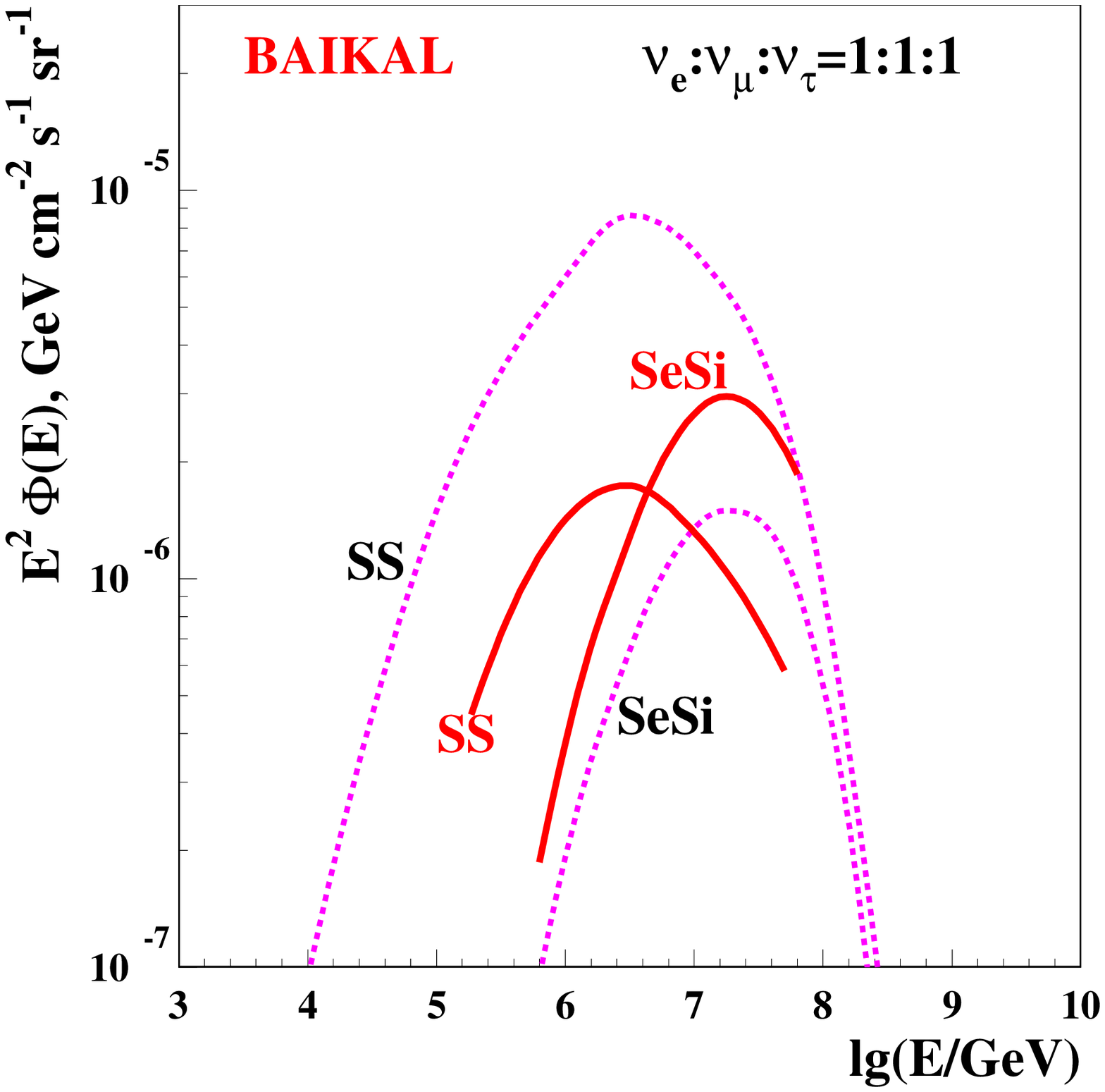}
\\
\caption{Left panel: all-flavor neutrino flux 
predictions in different models of neutrino sources 
compared to experimental upper limits to $E^{-2}$ fluxes obtained
by this analysis and other experiments (see text).
Right panel: Baikal
experimental limits compared to two
model predictions. Dotted curves: predictions
from model SS \cite{SS} and SeSi \cite{SeSi}. Full curves:
upper limits to spectra of the same shape.
Model SS is excluded (MRF=0.25), model SeSi is not
(MRF=2.12).
}
\label{fig16}
\end{figure*}
For an $E^{-2}$ behaviour of the neutrino spectrum and a flavor ratio 
$\nu_e:\nu_{\mu}:\nu_{\tau}=1:1:1$, the 90\% C.L. upper limit on the 
neutrino flux of all flavors obtained with the Baikal neutrino telescope  
{\it NT200} (1038 days) is:
$$
E^2\Phi<8.1 \times 10^{-7} 
\mbox{cm}^{-2}\mbox{s}^{-1}\mbox{sr}^{-1}\mbox{GeV}.
$$
\begin{equation}
\label{eq2}
\end{equation}

For the resonant process 

\begin{equation}
\bar{\nu_e} + e^- \rightarrow W^- \rightarrow \mbox{anything},
\end{equation}

\noindent
with the resonant neutrino energy  
$E_0=M^{2}_w/2m_e=6.3\times 10^6 \,$GeV 
and a cross section $5.02\times 10^{-31}$cm$^2$,
the event number is given by:
$$
N_{\bar{\nu_e}}=T \int d\Omega
\int dEV_{eff}\int\limits_{E^-}^{E^+}dE_{\nu}\times
$$
\begin{equation}
\label{eq4}
\times\Phi_{\bar{\nu_e}}(E_{\nu}) 
\frac{10}{18}N_{A} \rho_{H_2O} \frac{d\sigma_{\bar{\nu_e},e}}{dE},
\end{equation}
where $E^{\pm}=(M_w \pm 2\Gamma_w)^2/2m_e$ and
$M_w=$80.22 GeV, $\Gamma_w=$2.08 GeV.

Assuming an upper limit on the number of signal events $n_{90\%}=$2.5,
the model-independent limit on $\bar{\nu_e}$ at the W - resonance energy is: 
$$
\Phi_{\bar{\nu_e}} < 3.3 \times 10^{-20}
\mbox{cm}^{-2}\mbox{s}^{-1}\mbox{sr}^{-1}\mbox{GeV}^{-1}.
$$
\begin{equation}
\label{eq3}
\end{equation}

Figure \ref{fig16} (left panel) shows our upper limit on 
the all flavor $E^{-2}$ diffuse flux (\ref{eq2})
as well as the model independent limit on the resonant $\bar{\nu}_e$ flux 
(diamond) (\ref{eq3}). Also shown are the limits obtained by AMANDA and MACRO 
\cite{AMANDAHE,AMANDAMU,MACROHE}, theoretical bounds obtained by 
Berezinsky (model independent (B) and for an $E^{-2}$ shape
of the neutrino spectrum (B($E^{-2}$)) 
\cite{Ber3}, by Waxman and Bahcall (WB) \cite{WB1}, by Mannheim et al.(MPR) 
\cite{MPR}, predictions for neutrino fluxes from topological defects (TD) 
\cite{SeSi}, prediction on diffuse flux from AGNs according to Nellen et al. 
(NMB) \cite{NMB}, 
as well as the atmospheric 
conventional neutrino \mbox{fluxes \cite{VOL}} from horizontal and vertical 
directions ( ($\nu$) upper and lower curves, respectively) and atmospheric prompt 
neutrino fluxes ($\nu_{pr}$) obtained by Volkova et al. \cite{VPPROMPT}.
The right panel of fig. \ref{fig16} shows our upper limits (solid curves) on 
diffuse fluxes from AGNs shaped according to the model of Stecker and 
Salamon (SS) \cite{SS} and of Semikoz and Sigl (SeSi) \cite{SeSi}, 
according to Table 4.

\begin{figure}[htb]
\begin{center}
\includegraphics*[width=.45\textwidth,height=7.0cm]{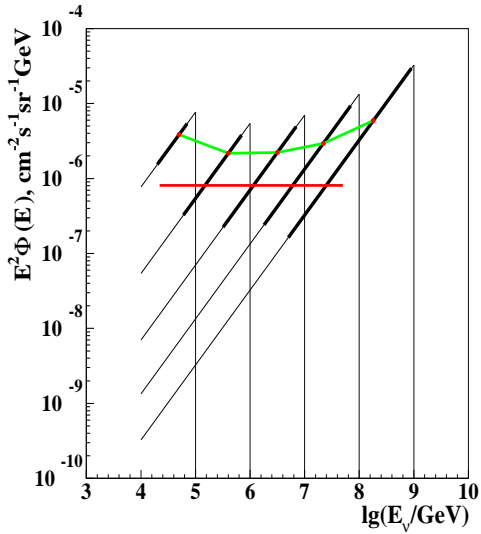}
\end{center}
\caption{Upper limits on $E^{-1}$ benchmark spectra with variable cutoff.
Thick lines cover the energy ranges containing
90\% of expected events. Points at the median energies
are connected by the thick curve. Also shown is the 
upper limit on $E^{-2}$ spectrum.}
\label{fig17}
\end{figure}

The diffuse neutrino flux is assumed to be composed of
contributions from cosmological
sources with different luminosity and energy spectra.
To benchmark the energy dependence of our limit
without referring to a special model, we show in
Fig.\ref{fig17} the upper limits derived for $E^{-1}$ spectra with 
variable, sharp cutoff. The parts of the spectra
drawn with thick lines cover the energy 
ranges containing 90\% of expected events.
Also shown is a curve connecting the points
which correspond to the median energies of recorded events.
This curve represents the upper limit
on a diffuse flux formed by superposition of the
benchmark spectra. Also shown (horizontal line) 
is the upper limit on an $E^{-2}$ spectrum, as given in (\ref{eq2}) above.

\section{Conclusion}
The neutrino telescope {\it NT200} in Lake Baikal is 
taking data since April 1998. Due to high water transparency and 
low light scattering, the detection volume of {\it NT200} for high 
energy $\nu_e$, $\nu_{\mu}$ and $\nu_{\tau}$ events
is several megatons and
exceeds the geometrical volume by a factor of about 50 for highest
energies. This results in a high sensitivity to diffuse 
neutrino fluxes from extraterrestrial sources -- more than an
order of magnitude better than that of underground searches and
similar to the published limits of AMANDA, 
the other operating large neutrino
telescope. The upper limit obtained for a diffuse 
($\nu_e+\nu_{\mu}+\nu_{\tau}$) flux with $E^{-2}$ shape
is $E^2 \Phi = 8.1 \times 10^{-7}$cm$^{-2}$s$^{-1}$sr$^{-1}$GeV.
With  $3.3 \times 10^{-20}$cm$^{-2}$s$^{-1}$sr$^{-1}$GeV$^{-1}$,
the limit on a $\bar{\nu_e}$ flux
at the resonant  energy 6.3$\times$10$^6$GeV is presently the most
stringent.

To extend the search for diffuse extraterrestrial neutrinos 
with higher sensitivity, {\it NT200} was significantly upgraded.
\cite{NT+}.
In March/April 2005 we fenced a large part of the
search volume (see Fig.12, right) with three sparsely
instrumented strings. The three-year sensitivity of this
enlarged detector {\it NT200}$+$ on the neutrino flux of all flavors, 
with about 5 Mton enclosed volume, is approximately 
$2\times10^{-7}$cm$^{-2}$s$^{-1}$sr$^{-1}$GeV
for $E>$10$^2$ TeV, i.e. three-four times better than that of {\it NT200}.
{\it NT200}$+$ will search for neutrinos from AGNs, GRBs
and other extraterrestrial sources, neutrinos from cosmic ray
interactions in the Galaxy as well as high energy atmospheric muons
with $E_{\mu}>10$ TeV.

\section{Acknowledgments}
{\it This work was supported by the Russian Ministry of Education and Science,
the German Ministry of Education and Research and the Russian Fund of Basic 
Research} ({\it grants} \mbox{\sf 05-02-17476} {\it and} 
\mbox{\sf 04-02-17289}), {\it and by the Grant of President of 
Russia} \mbox{\sf NSh-1828.2003.2}. 

\end{document}